\pdfoutput=1
\documentclass[useAMS,usenatbib]{openjournal}
\usepackage{times}
\usepackage{graphicx}
\usepackage{subfigure}
\usepackage{psfrag}
\usepackage{amsmath}
\usepackage{amssymb}
\usepackage{natbib}
\usepackage{bm}

\usepackage[title]{appendix}
\usepackage{epstopdf}
\usepackage[utf8]{inputenc}

\DeclareUnicodeCharacter{2212}{-} 

\def\reff@jnl#1{{\rm#1\/}}
\def\aj{\reff@jnl{AJ}}                 
\def\araa{\reff@jnl{ARA\&A}}           
\def\apj{\reff@jnl{ApJ}}               
\def\apjl{\reff@jnl{ApJ}}              
\def\apjs{\reff@jnl{ApJS}}             
\def\ao{\reff@jnl{Appl.Optics}}        
\def\apss{\reff@jnl{Ap\&SS}}           
\def\aap{\reff@jnl{A\&A}}              
\def\aapr{\reff@jnl{A\&A~Rev.}}        
\def\aaps{\reff@jnl{A\&AS}}            
\def\azh{\reff@jnl{AZh}}               
\def\baas{\reff@jnl{BAAS}}             
\def\jcap{\reff@jnl{JCAP}}             
\def\jrasc{\reff@jnl{JRASC}}           
\def\memras{\reff@jnl{MmRAS}}          
\def\mnras{\reff@jnl{MNRAS}}           
\def\pra{\reff@jnl{Phys.Rev.A}}        
\def\prb{\reff@jnl{Phys.Rev.B}}        
\def\prc{\reff@jnl{Phys.Rev.C}}        
\def\prd{\reff@jnl{Phys.Rev.D}}        
\def\prl{\reff@jnl{Phys.Rev.Lett}}     
\def\pasp{\reff@jnl{PASP}}             
\def\pasj{\reff@jnl{PASJ}}             
\def\qjras{\reff@jnl{QJRAS}}           
\def\skytel{\reff@jnl{S\&T}}           
\def\solphys{\reff@jnl{Solar~Phys.}}   
\def\sovast{\reff@jnl{Soviet~Ast.}}    
\def\ssr{\reff@jnl{Space~Sci.Rev.}}    
\def\zap{\reff@jnl{ZAp}}               
\def\nat{\reff@jnl{Nature}}            

\begin{document}

\title{Importance Nested Sampling and the {\sc MultiNest} Algorithm}

\author{F.~Feroz$ ^1 $\thanks{E-mail: f.feroz@mrao.cam.ac.uk}, M.P.~Hobson$ ^1 $, E.~Cameron$ ^2 $ and A.N.~Pettitt$ ^3$\\
$^1$Astrophysics Group, Cavendish Laboratory, JJ Thomson Avenue, Cambridge, UK\\
$^2$Big Data Institute, Li Ka Shing Centre for Health Information and Discovery, University of Oxford, UK\\
$^3$School of Mathematical Sciences (Statistical Science), Queensland University of Technology (QUT), Brisbane, Australia\\}

\label{firstpage}

\maketitle

Bayesian inference involves two main computational challenges. First,
in estimating the parameters of some model for the data, the posterior
distribution may well be highly multi-modal: a regime in which the
convergence to stationarity of traditional Markov Chain Monte Carlo
(MCMC) techniques becomes incredibly slow. Second, in selecting
between a set of competing models the necessary estimation of the
Bayesian evidence for each is, by definition, a (possibly
high-dimensional) integration over the entire parameter space; again
this can be a daunting computational task, although new Monte Carlo
(MC) integration algorithms offer solutions of ever increasing
efficiency.  Nested sampling (NS) is one such contemporary MC strategy
targeted at calculation of the Bayesian evidence, but which also
enables posterior inference as a by-product, thereby allowing
simultaneous parameter estimation and model selection. The widely-used
{\sc MultiNest} algorithm presents a particularly efficient
implementation of the NS technique for multi-modal posteriors. In this
paper we discuss importance nested sampling (INS), an alternative
summation of the {\sc MultiNest} draws, which can calculate the
Bayesian evidence at up to an order of magnitude higher accuracy than
`vanilla' NS with no change in the way {\sc MultiNest} explores the
parameter space. This is accomplished by treating as a
(pseudo-)importance sample the totality of points collected by {\sc
MultiNest}, including those previously discarded under the constrained
likelihood sampling of the NS algorithm. We apply this technique to
several challenging test problems and compare the accuracy of Bayesian
evidences obtained with INS against those from vanilla NS.



\section{Introduction}\label{sec:intro}

The last two decades in astrophysics and cosmology have seen the arrival of vast amounts of high quality data. To facilitate inference regarding the physical processes under investigation, Bayesian methods have become increasingly important and widely used (see e.g. \citealt{2008ConPh..49...71T} for a review). In such applications, the process of Bayesian inference may be sensibly divided into two distinct categories: parameter estimation and model selection. Parameter estimation is typically achieved via MCMC sampling methods based on the Metropolis-Hastings algorithm and its variants, such as slice and Gibbs sampling (see e.g. \citealt{MacKay}). Unfortunately, these methods can be highly inefficient in exploring multi-modal or degenerate distributions. Moreover, in order to perform Bayesian model selection \citep{clyde}, estimation of the Bayesian `evidence', or marginal likelihood, is needed, requiring a multi-dimensional integration over the prior density. Consequently, the computational expense involved in Bayesian model selection is typically an order of magnitude greater than that for parameter estimation, which has undoubtedly hindered its use in cosmology and astroparticle physics to-date.

Nested sampling (NS; \citealt{skilling04, ski06, Sivia}) is a contemporary Monte Carlo (MC) method targeted at the efficient calculation of the evidence, yet which allows posterior inference as a by-product, providing a means to carry out simultaneous parameter estimation and model selection (and, where appropriate, model averaging). \citet{feroz08} and \citet{multinest} have built on the NS framework by introducing the now-popular {\sc MultiNest} algorithm, which is especially efficient in sampling from posteriors that may contain multiple modes and/or degeneracies. This technique has already greatly reduced the computational cost of Bayesian parameter estimation and model selection and has successfully been applied to numerous inference problems in astrophysics, cosmology and astroparticle physics (see e.g. \citealt{2009CQGra..26u5003F, 2010CQGra..27g5010F, 2011MNRAS.416L.104F, 2011MNRAS.415.3462F, 2009MNRAS.400.1075B, 2012MNRAS.421..169G, 2010JHEP...07..064W, 2012ApJ...750..115K, 2012arXiv1207.3708K, 2013MNRAS.429.1278K, 2012arXiv1212.2636S, Teacheyeaav1784}).

In this paper, we discuss importance nested sampling (INS), an alternative summation of the draws from {\sc MultiNest}'s exploration of the model parameter space with the potential to increase its efficiency in evidence computation by up to a order-of-magnitude. Version (v3.0) of {\sc MultiNest}, which implements INS in addition to the vanilla NS scheme of previous versions, is available at {\tt https://github.com/farhanferoz/MultiNest}.

The outline of this paper is as follows. We give a brief introduction to Bayesian inference in Sec.~\ref{sec:bayesian} and describe nested sampling along with the {\sc MultiNest} algorithm in Sec.~\ref{sec:multinest}. The INS technique is discussed in Sec.~\ref{sec:INS} and is applied to several test problems in Sec.~\ref{sec:applications}. We summarize our findings in Sec.~\ref{sec:discussion}. Finally, in Appendix A we discuss the relationship between INS and other contemporary MC schemes, in Appendix B we give a detailed account of the convergence properties of INS within the {\sc MultiNest} algorithm, and in Appendix C we present a brief measure-theoretic commentary on vanilla NS.

\section{Bayesian Inference}\label{sec:bayesian}

Bayesian inference provides a principled approach to the inference of a
set of parameters, $\mathbf{\Theta}$, in a model (or hypothesis), $H$,
for data, $\mathbf{D}$. Bayes' theorem states that
\begin{equation} 
\Pr(\mathbf{\Theta}|\mathbf{D}, H) = \frac{\Pr(\mathbf{D}|\,\mathbf{\Theta},H)\Pr(\mathbf{\Theta}|H)}{\Pr(\mathbf{D}|H)},
\end{equation}
where $\Pr(\mathbf{\Theta}|\mathbf{D}, H) \equiv P(\mathbf{\Theta}|\mathbf{D})$ is the posterior probability
density of the model parameters, $\Pr(\mathbf{D}|\mathbf{\Theta}, H)
\equiv \mathcal{L}(\mathbf{\Theta})$ the likelihood of the data, and
$\Pr(\mathbf{\Theta}|H) \equiv \pi(\mathbf{\Theta})$ the parameter prior. The
final term, $\Pr(\mathbf{D}|H) \equiv \mathcal{Z}$ (the Bayesian
evidence), represents the factor required to normalize the posterior
over the domain of $\mathbf{\Theta}$ given by:
\begin{equation}
\mathcal{Z} = \int_{\Omega_{\bm{\Theta}}}{\mathcal{L}(\mathbf{\Theta})\pi(\mathbf{\Theta})}d\mathbf{\Theta}.
\label{eq:Z}
\end{equation} 
Being independent
of the parameters, however, this factor can be ignored in
parameter inference problems  which can be approximated by taking
samples from the unnormalized posterior only, using standard MCMC
methods (for instance).

Model selection between two competing models, $H_{0}$ and $H_{1}$, can be achieved by comparing their respective posterior probabilities given the observed dataset as follows:
\begin{equation}
R = \frac{\Pr(H_{1}|\mathbf{D})}{\Pr(H_{0}|\mathbf{D})}
  = \frac{\Pr(\mathbf{D}|H_{1})\Pr(H_{1})}{\Pr(\mathbf{D}| H_{0})\Pr(H_{0})}
  = \frac{\mathcal{Z}_1}{\mathcal{Z}_0} \frac{\Pr(H_{1})}{\Pr(H_{0})}.
\label{eq:R}
\end{equation}
Here $\Pr(H_{1})/\Pr(H_{0})$ is the prior probability ratio for the
two models, which can often be set to unity in situations where there
is no strong \textit{a priori} reason for preferring one model over the
other, but occasionally requires further consideration (as in the
 Prosecutor's Fallacy; see also
\citealt{2008arXiv0810.0781F,2009MNRAS.398.2049F} for key
astrophysical examples). It can be seen from Eq.~(\ref{eq:R}) that the Bayesian
evidence thus plays a central role in Bayesian model selection.

As the average of the likelihood over the prior, the evidence is
generally larger for a model if more of its parameter space is likely and
smaller for a model with large areas in its parameter space having low
likelihood values, even if the likelihood function is sharply
peaked. Thus, the evidence may be seen both as penalizing `fine tuning' of
a model against the observed data and as an automatic implementation of Occam's Razor.

\section{Nested Sampling and the {\sc MultiNest} Algorithm}\label{sec:multinest}

%
\begin{figure}
\begin{center}
\subfigure[]{\includegraphics[width=0.3\columnwidth,height=0.3\columnwidth]{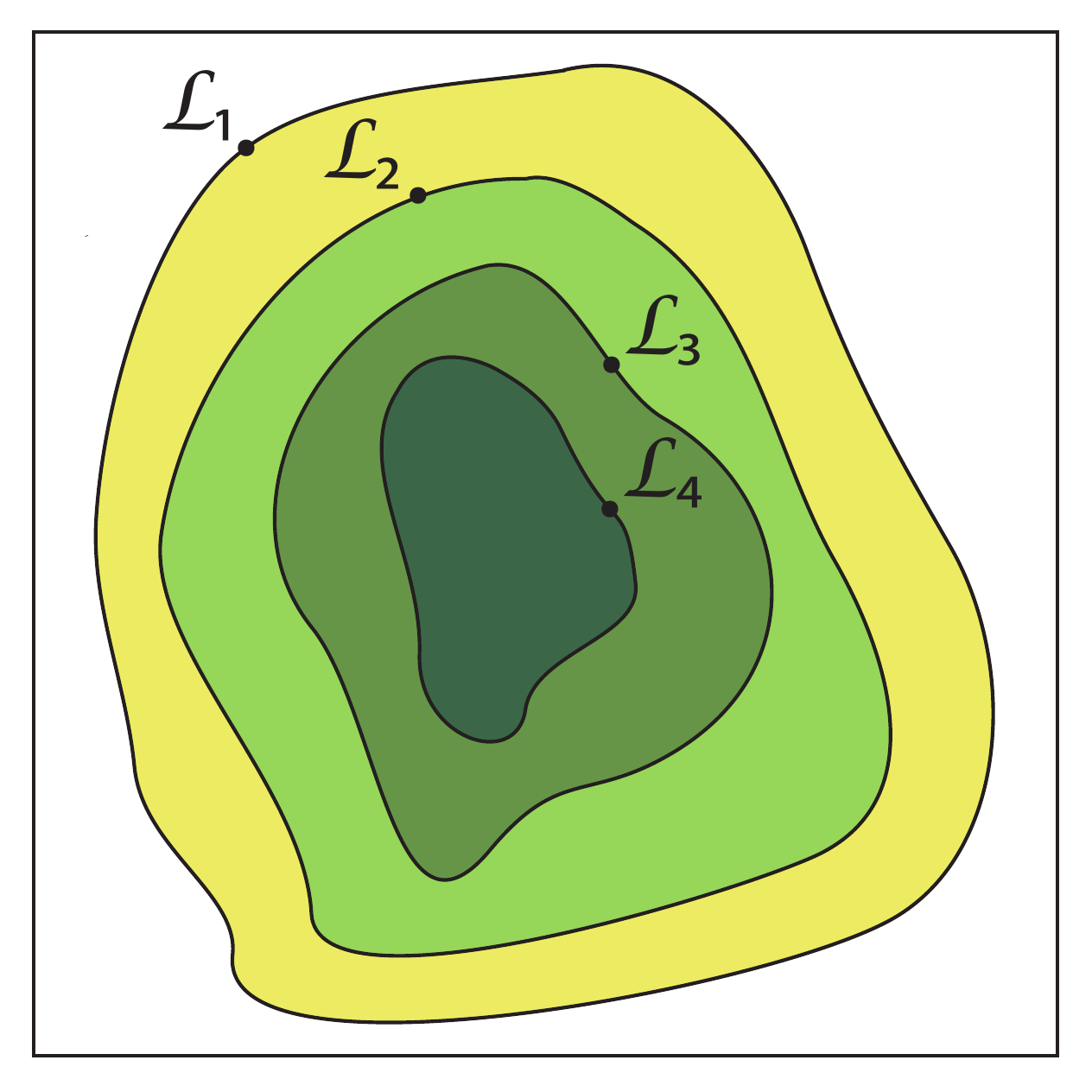}}\hspace{0.3cm}
\subfigure[]{\includegraphics[width=0.3\columnwidth,height=0.3\columnwidth]{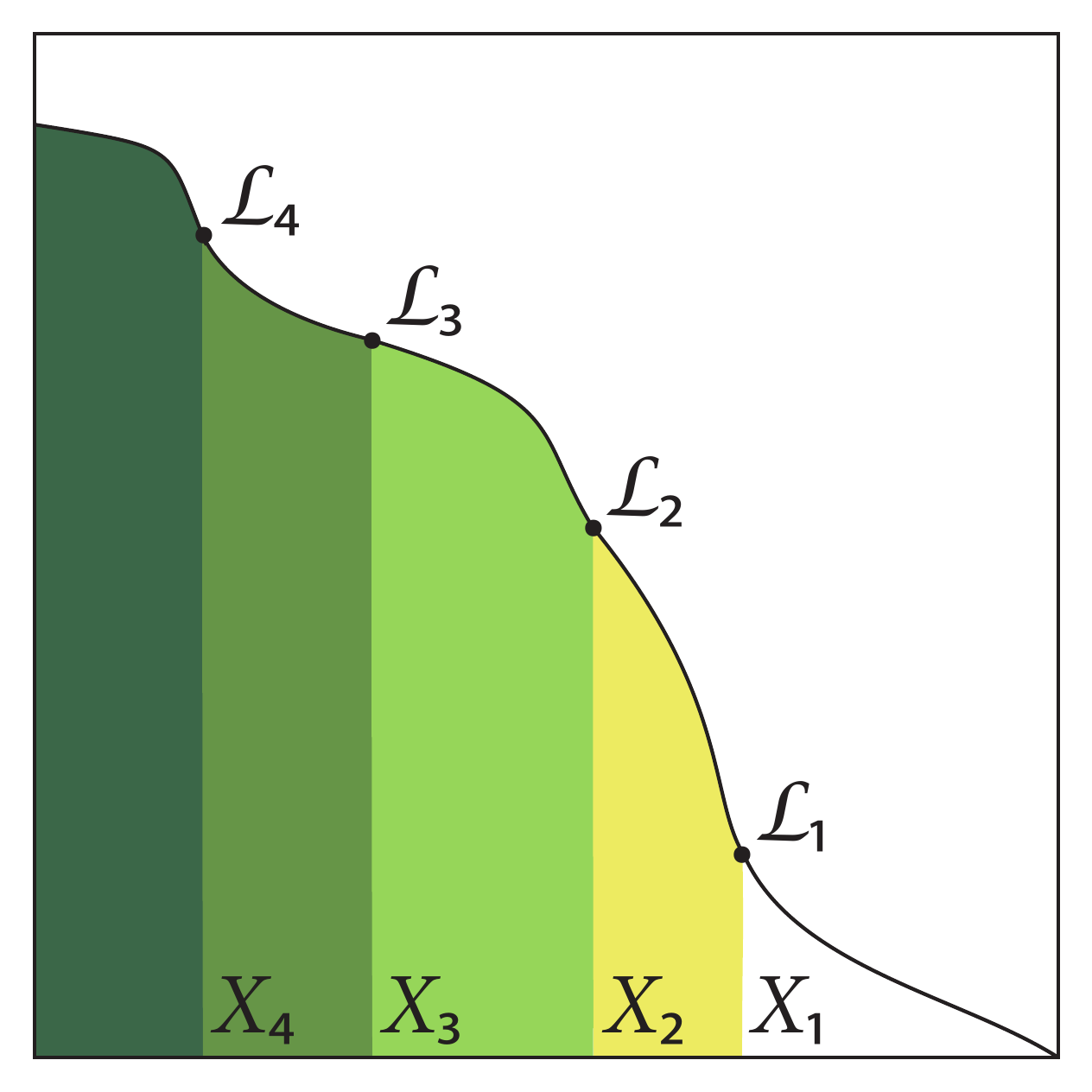}}
\caption{Cartoon illustrating (a) the posterior of a two dimensional problem; and (b) the transformed
$\mathcal{L}(X)$ function where the prior volumes, $X_{i}$, are associated with each likelihood, $\mathcal{L}_{i}$.}
\label{fig:NS}
\end{center}
\end{figure}

Nested sampling estimates the Bayesian evidence by transforming
the multi-dimensional evidence integral over the prior density into a
one-dimensional integral over an inverse survival function (with
respect to prior mass) for the likelihood itself. This is accomplished
by considering  the survival function, $X(\lambda)$, for
$\mathcal{L}(\mathbf{\Theta})$, dubbed ``the prior volume'' here; 
namely, 
\begin{equation}
X(\lambda) = \int_{\{\bm{\Theta} : \mathcal{L}\left(\mathbf{\Theta}\right) > \lambda\}}
\pi(\mathbf{\Theta}) d \mathbf{\Theta},
\label{eq:Xdef}
\end{equation}
where the integral extends over the region(s) of parameter space
contained within the iso-likelihood contour,
$\mathcal{L}(\mathbf{\Theta}) = \lambda$.  Recalling that the
expectation value of a non-negative random variable may be recovered
by integration over its survival function (a result evident from
integration by parts) we have (unconditionally):
\begin{equation}
\mathcal{Z}=\int_{0}^{\infty}X(\lambda)d\lambda.
\end{equation}
 When $\mathcal{L}(X)$, the inverse of $X(\lambda)$, exists (i.e., when $\mathcal{L}(\mathbf{\Theta})$ is a continuous function with connected
support; \citealt{cho10}) the evidence integral may thus be further
rearranged as:
\begin{equation}
\mathcal{Z}=\int_{0}^{1}{\mathcal{L}(X)}dX.
\label{eq:nested}
\end{equation}
Indeed, if $\mathcal{L}(X)$ were 
known exactly (and Riemann integrable\footnote{We give a brief
  measure-theoretic formulation of NS in Appendix C.}), by evaluating the likelihoods,
$\mathcal{L}_{i}=\mathcal{L}(X_{i})$, for a deterministic sequence of
$X$ values,
\begin{equation}
0<X_{N}<\cdots <X_{2}<X_{1}< X_{0}=1,
\end{equation}
as shown schematically in Fig.~\ref{fig:NS}, the evidence could in
principle be approximated numerically using only standard quadrature methods as follows:
\begin{equation}
\mathcal{Z} \approx \hat{\mathcal{Z}}={\textstyle {\displaystyle \sum_{i=1}^{N}}\mathcal{L}_{i}w_{i}},
\label{eq:NS_sum}
\end{equation}
where the weights, $w_{i}$, for the simple trapezium rule are given by
$w_{i}=\frac{1}{2}(X_{i-1}-X_{i+1})$.  With $\mathcal{L}(X)$ typically
unknown, however, we must turn to MC methods for the probabilistic
association of prior volumes, $X_i$, with likelihood contours,
$\mathcal{L}_i=\mathcal{L}(X_i)$, in our computational evidence estimation.

\subsection{Evidence estimation}\label{app:nested:evidence}

Under the default nested sampling algorithm the summation in
Eq.~(\ref{eq:NS_sum}) is performed as follows. First $N_{\rm live}$
`live' points are drawn from the prior,
$\pi(\mathbf{\Theta})$, and the initial prior volume, $X_{0}$, is set
to unity. At each subsequent iteration, $i$, the point with lowest
likelihood value, $\mathcal{L}_{i}$, is removed from the live point
set and replaced by another point drawn from the prior 
under the constraint that its likelihood is higher than
$\mathcal{L}_{i}$.  The prior volume contained within this region at the
$i^{\rm th}$ iteration, is thus a random variable distributed as $X_{i} = t_{i}
X_{i-1}$, where $t_{i}$ follows the distribution for the
largest of $N_{\rm live}$ samples drawn uniformly from the interval
$[0,1]$ (i.e., $\Pr(t) = N_{\rm
  live}t^{N_{\rm live}-1}$). This sampling process is repeated until (effectively) the entire prior
volume has been traversed; the live particles moving through \textit{nested}
shells of constrained likelihood as the prior volume is steadily
reduced. The mean and standard deviation of $\log t$, which governs
the geometrical exploration of the prior volume, are:
\begin{equation}
E[\log t] = -1/N_{\rm live}, \quad \sigma[\log t] = 1/N_{\rm live}.
\end{equation}
Since each draw of $\log t_i$ is independent here, after $i$ iterations the
prior volume will shrink down as $\log X_{i} \approx
-(i\pm\sqrt{i})/N_{\rm live}$. Thus, one may take $X_{i} \approx \exp(-i/N_{\rm live})$.

\subsection{Stopping criterion}\label{nested:stopping}

The NS algorithm should terminate when the expected evidence
contribution from the current set of live points is less than a
user-defined tolerance. This expected remaining contribution can be
estimated (cautiously) as $\Delta{\mathcal{Z}}_i =
\mathcal{L}_{\rm max}X_i$, where ${\cal L}_{\rm max}$ is the
maximum likelihood value amongst the current set of live points (with $X_i$
the expected value of remaining prior volume, as before).

\subsection{Posterior inferences}\label{nested:posterior}

Although the NS algorithm is designed specifically to estimate the
Bayesian evidence, inferences from the posterior distribution can be
easily obtained using the final live points and the full sequence of
discarded points from the NS process, i.e., the points with the
lowest likelihood value at each iteration of the algorithm. Each
such point is simply assigned the importance weight,
\begin{equation}
p_{i} = \frac{\mathcal{L}_{i}w_{i}}{\sum_j \mathcal{L}_{j}w_{j}} =\frac{\mathcal{L}_{i}w_{i}}{\hat{\mathcal{Z}}},
\label{eq:12}
\end{equation}
from which sample-based
estimates for the key posterior parameter summaries (e.g.\ means, standard
deviations, covariances and so on) may be computed\footnote{Some
  relevant commentary on this aspect of NS with regard to Lemma 1 of
  \citet{cho10} appears in Appendix C.}. (As a
self-normalizing importance sampling estimator the asymptotic variance of these
moments is of course dependent upon both the similarity between the NS
path and the target and the accuracy of $\hat{\mathcal{Z}}$
itself; cf.\ \citealt{hes95}.) 
Readers unfamiliar with importance sampling (IS) ideas may refer to \citet{liu}
for an insightful overview of this topic and its application to
diverse branches of modern science (including statistical physics,
cell biology, target tracking, and genetic analysis).

\subsection{Practical implementations of nested sampling}\label{nested:NS_variants}

The main challenge in implementing the computational NS algorithm is to draw unbiased samples efficiently from the likelihood-constrained prior. John Skilling, originally proposed to use the Markov Chain Monte Carlo (MCMC) method for this purpose (\citealt{skilling04, ski06, Sivia}). One such implementation (\citealt{2010PhRvD..81f2003V}), with specific proposal distributions for the MCMC step, has been used successively in gravitational wave searches.

In astrophysics in particular, rejection sampling schemes have been successfully employed to draw samples from the likelihood-constrained prior. It was first proposed in the {\sc CosmoNest} package (\citealt{muk06}) through the use of ellipsoidal rejection sampling scheme and was shown to work very well for uni-modal posterior distributions. This method was improved upon in {\sc CosmoClust} package (\citealt{cosmoclust}) through the use of a clustering scheme to deal with multi-modal distributions. {\sc MultiNest} was then proposed with several innovations to make ellipsoidal rejection sampling more robust in dealing with multi-modal distributions. Other methods employing ellipoidal rejection sampling scheme within Nested Sampling framework include the {\sc Diamonds} (\citealt{diamonds}) and {\sc Dynesty} (\citealt{dynesty}) packages.

One particular problem with rejection sampling schemes is the exponential reduction in sampling efficiency with increasing dimensionality of the problem. In order to address this issue, a slice sampling method has been employed to draw unbiased samples efficiently from the likelihood-constrained prior in the {\sc PolyChord} (\citealt{polychorda, polychordb}) package.

Another algorithm to increase the efficiency of Nested Sampling through the variable number of live points is the ``Dynamic Nested Sampling" method (\citealt{higson18}) which has been used in the {\sc Dynesty} (\citealt{dynesty}) package.

\subsection{{\sc MultiNest} algorithm}\label{sec:method:bayesian:multinest}
%
\begin{figure}
\begin{center}
\subfigure[]{\fbox{\includegraphics[width=0.3\columnwidth,height=0.3\columnwidth]{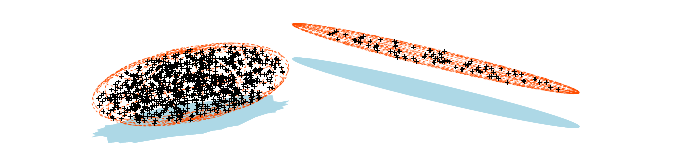}}}
\setcounter{subfigure}{1}
\subfigure[]{\fbox{\includegraphics[width=0.3\columnwidth,height=0.3\columnwidth]{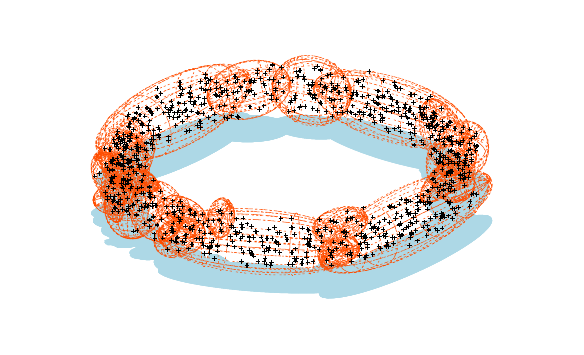}}}
\setcounter{subfigure}{2}
\caption{Illustrations of the ellipsoidal decompositions returned by {\sc MultiNest}. The points given as input
are overlaid on the resulting ellipsoids. Here 1000 points were sampled uniformly from: (a) two non-intersecting
ellipsoids; and (b) a torus.}
\label{fig:dino}
\end{center}
\end{figure}

The {\sc MultiNest} algorithm \citep{feroz08, multinest} addresses this
problem of drawing unbiased samples from the likelihood-constrained prior, through an ellipsoidal rejection sampling scheme. At each iteration,
$i$, the full set of $N_{\rm live}$ live points is enclosed within a
set of (possibly overlapping) ellipsoids and the desired replacement
point sought
 from within their union. The ellipsoidal decomposition of the live
 point set is performed through an expectation-minimisation algorithm
 such that the sum of volumes of the ellipsoids is minimised with the additional
 constraint  that the
 total volume enclosed by the ellipsoids is at least  
 $X_{i}/f$.  Again $X_{i} \approx \exp(-i/N_{\rm live})$ is the
 expected prior volume, while $0 < f \le 1$ is a user defined value
 for the target efficiency (the ratio of points accepted to points
 sampled).  Thus, $f$ is analogous to the (inverse of the)
 ``enlargement factor'' introduced by \citet{muk06} into their
 pioneering ellipsoid-based NS code; the larger the target $f$ the
 faster the algorithm runs, but the greater the chance of some
 ellipsoids failing to cover the full $\mathcal{L}>\mathcal{L}_i$
 volume (biasing the vanilla NS estimates, though not
 necessarily the INS estimates, as we discuss later).

The {\sc MultiNest} ellipsoidal decomposition algorithm thus allows
substantial flexibility in the geometry of its posterior exploration; with bent and/or
irregularly-shaped posterior modes typically broken into a relatively large
number of small ‘overlapping’ ellipsoids and smooth, near-Gaussian
posterior modes kept whole (or broken into relatively few ellipsoids),
as shown in Fig.~\ref{fig:dino}. It thereby automatically
accommodates elongated, curving degeneracies while maintaining high
efficiency for simpler problems. {\sc MultiNest} also specifically enables the identification of
distinct modes by isolating non-overlapping subsets of the ellipsoidal
decomposition; so identified, these distinct modes can then be evolved
independently.

Once the ellipsoidal bounds have been created at a given iteration of
the {\sc MultiNest} algorithm a new point is drawn uniformly from
the union of these ellipsoids as follows. If there are $L$ ellipsoids
 at iteration $i$, a particular ellipsoid is chosen with probability $p_{l}$ given as:
\begin{equation}
p_{l} = V_{l} / V_{\rm tot},
\end{equation}
where $V_{\rm tot} = \sum_{l=1}^{L} V_{l}$, from which a single point is then drawn
uniformly and checked against
the constraint $\mathcal{L}> \mathcal{L}_i$. If satisfied the point
is accepted with probability $1/q$, where $q$ is the number of
ellipsoids the new point lies in (in order to take into account the
possibility of non-empty intersections), otherwise it is rejected (but
saved for INS summation) and the process is repeated with a new random
choice of ellipsoid.

In higher dimensions, most of the volume of an ellipsoid lies in its outer shells and therefore any overshoot of the  ellipsoidal decomposition relative to the true iso-likelihood surface can result in a marked drop in sampling efficiency. In order to maintain the sampling efficiency for such high dimensional problems, {\sc MultiNest} can also operate in a `constant efficiency mode'. In this mode, the total volume enclosed by the ellipsoids is no longer linked with the expected prior volume $X_{i}$ by requiring the total ellipsoidal volume to be at least $X_{i}/f$, instead the total volume enclosed by the union of ellipsoids is adjusted such that the sampling efficiency is as close to the user defined target efficiency $f$ as possible while keeping every live point enclosed in at least one ellipsoid. Despite the increased chance of the fitted ellipsoids encroaching \textit{within} the constrained-likelihood volume (i.e., missing regions of parameter space for which $\mathcal{L}> \mathcal{L}_{i}$), past experience has shown (e.g.\ \citealt{multinest}) this constant efficiency mode may nevertheless produce reasonably accurate posterior distributions for parameter estimation purposes. The vanilla NS evidence values, however, cannot be relied upon in this mode, with the expectation being a systematic over-estimation of the model evidence. Interestingly, the same is not \textit{strictly} true of the INS evidence estimates, which use the NS technique only for posterior exploration (not evidence summation); though we will note later some important caveats for its error estimation.

In the rest of this paper, we refer to the mode in which {\sc MultiNest} links the volume of the ellipsoidal decomposition with the expected prior volume as its `default' mode, and we specifically highlight instances where `constant efficiency mode' has been trialled.

\section{Importance Nested Sampling}\label{sec:INS}

Though highly efficient in its approximation to the iso-likelihood
contour bounding the live particle set at each iteration, the
ellipsoidal rejection sampling scheme used by {\sc MultiNest}
ultimately discards a significant pool of sampled points failing to
satisfy the NS constraint, $\mathcal{L}> \mathcal{L}_i$, for which the
likelihood has nevertheless been evaluated at some computational cost.
In order to redress this final inefficiency the technique of
importance nested sampling (INS) has recently been proposed
by \cite{cam13} as an alternative summation for the
Bayesian evidence in this context. In particular, INS uses all the
points drawn by {\sc MultiNest}, or any other ellipsoidal rejection
sampling algorithm, at each iteration regardless of whether they
satisfy the constraint $\mathcal{L}> \mathcal{L}_i$ or not.  The
relationship of INS to existing MC schemes is summarised in
Appendix~A.

\subsection{Pseudo-importance sampling density}

One begins by defining the following pseudo-importance sampling
density:
\begin{equation} 
g(\mathbf{\Theta}) = \frac{1}{N_{\mathrm{tot}}}\sum_{i=1}^{N_{\mathrm{iter}}} \frac{n_{i} E_{i}(\mathbf{\Theta})}{V_{{\rm tot},i}},
\label{eq:IS-density}
\end{equation}
where $N_{\mathrm{iter}}$ is the total number of iterations (ellipsoidal
decompositions) performed by {\sc MultiNest}, $n_{i}$ the number of
points collected at the $i^\mathrm{th}$ iteration (with total, $N_{\mathrm{tot}} =
\sum_{i=1}^{N_{\mathrm{iter}}} n_{i}$), $V_{{\rm tot},i}$ the total volume enclosed in the 
union of ellipsoids at the $i^\mathrm{th}$ iteration, and
$E_{i}(\mathbf{\Theta})$ an indicator function returning $1$ when
$\mathbf{\Theta}$ lies in the $i^\mathrm{th}$ ellipsoidal
decomposition and $0$ otherwise.  We call $g(\mathbf{\Theta})$ here a
\textit{pseudo}-importance sampling density since it is of course
defined only
\textit{a posteriori} to our sampling from it, with the consequence
that all $\mathbf{\Theta} \sim E_{j>i}(\mathbf{\Theta})$ are to some
(ideally negligible) extent dependent on all previous $\mathbf{\Theta}
\sim E_{j\le i}(\mathbf{\Theta})$ (some important implications of which
we discuss in Sec.~\ref{INS:error} below).  The heritage of this
technique lies with the reverse logistic regression strategy of
\citet{gey94} and the ``biased sampling'' framework of \citet{var85}.
Another term that has been used in place of pseudo-importance sampling
is ``recycling of past draws'' (e.g.\ \citealt{cor12}).

If at each iteration, the ellipsoidal decomposition would consist of only one ellipsoid then $V_{{\rm tot},i}$
is simply the geometric volume of the ellipsoid at iteration $i$. {\sc
MultiNest}, however, may enclose its live points in a set of possibly
overlapping ellipsoids. An analytical expression for calculating the
volume in the overlapped region of ellipsoids is not available and
therefore we estimate the volume occupied by the union of ellipsoids
through the following MC method. Whenever an ellipsoidal
decomposition is constructed, we draw $M$ points ($\mathbf{\Theta}_m^\prime$,
$m=1,2,\ldots,M$) from it as follows: for each draw we first pick an
ellipsoid with probability $V_{l}/\sum_{l=1}^L V_{l}$, where $V_{l}$
are the volumes of the $L$ ellipsoids in the decomposition; a point
$\mathbf{\Theta}_m^\prime$ is then drawn uniformly from the chosen ellipsoid
and we calculate, $q_m$, the number of ellipsoids it lies in. The
volume in the union of ellipsoids is then:
\begin{equation} 
V_{\rm tot} \approx \hat{V}_{\rm tot} = \frac{M}{\sum_{m=1}^{M} q_{m}} 
\sum_{l=1}^{L} V_{l}.
\label{IS-V}
\end{equation}
We note that this Monte Carlo procedure does not require any
evaluations of the likelihood function, and thus is not
computationally demanding.

\subsection{Evidence estimation and posterior samples}

As an alternative to the canonical NS summation given by
Eq.~(\ref{eq:NS_sum}) the Bayesian evidence
can instead be estimated with reference to the above pseudo-importance sampling density as:
\begin{equation} 
\mathcal{\hat{Z}} = \frac{1}{N_{\mathrm{tot}}} \sum_{k=1}^{N_{\mathrm{tot}}} \frac{\mathcal{L}(\mathbf{\Theta}_{k}) \pi(\mathbf{\Theta}_{k})}{g(\mathbf{\Theta}_{k})}.
\label{IS-Z}
\end{equation}
Moreover, each one of the $N_{\mathrm{tot}}$ points
collected by {\sc MultiNest} can be assigned the following estimator
of its posterior
probability density:
\begin{equation} 
P(\mathbf{\Theta}) = \frac{\mathcal{L}(\mathbf{\Theta}) \pi(\mathbf{\Theta})}{N_{\mathrm{tot}} g(\mathbf{\Theta})}.
\label{IS-p}
\end{equation}
Since the importance nested sampling scheme does not rely on the
ellipsoidal decomposition fully enclosing the region(s) satisfying the
constraint $\mathcal{L}> \mathcal{L}_i$, it can also achieve accurate
evidence estimates and  posterior summaries from sampling done in the
constant efficiency mode of {\sc MultiNest}.\footnote{  The reasons for this
are described in detail in Appendix B; but in brief we note that like
`ordinary' importance sampling the only fundamental constraint on the
$g(\bm{\Theta})$ of the INS scheme is that its support enclose that of the
posterior, which we ensure by drawing our first set of points from the
prior support itself, $E_1(\bm{\Theta})=1$ whenever $\pi(\bm{\Theta})
> 0$.} However, as we discuss
shortly, the utility of this feature is often limited by ensuing
difficulties in the estimation of uncertainty for such constant
efficiency mode evidence estimates.

From a computational perspective we note that in a na\"ive application
of this scheme it will be necessary to store $N_{\mathrm{tot}}$
points, $\mathbf{\Theta}_{k}$, along with the likelihood,
$\mathcal{L}(\mathbf{\Theta}_{k})$, and prior probability,
$\pi(\mathbf{\Theta}_{k})$, for each, \textit{as well as} all relevant
information describing the ellipsoidal decompositions (centroids,
eigen-values and eigen-vectors) at each iteration.  Even with a
Cholesky factorization of the eigen-vectors, storing the latter may
easily result in excessive memory requirements. However, since in the
{\sc MultiNest} algorithm the prior volume, and consequently the
volume occupied by the bounding ellipsoids, shrinks at each subsequent
iteration one can confidently assume $E_{i}(\mathbf{\Theta}) = 1$ for
all points drawn at iterations $j>i$.  At a given iteration then, one
needs only to check if points collected from previous iterations lie
in the current ellipsoidal decomposition and add the contribution to
$g(\mathbf{\Theta})$ coming from the current iteration as given in
Eq.~(\ref{eq:IS-density}). This results in an enormous reduction in
memory requirements as information about the ellipsoidal decomposition
from previous iterations no longer needs to be stored.

At each iteration, {\sc MultiNest} draws points from the ellipsoidal
decomposition, but in order to take account of the volume in the
overlaps between ellipsoids, each point is accepted only with
probability $1/q$ where $q$ is the number of
ellipsoids in which the given point lies. Rather than discarding all
these rejected points, which would be wasteful, we include them by
dividing the importance sampling weights as given in
Eq.~(\ref{eq:IS-density}), in three components:
\begin{equation} 
g(\mathbf{\Theta}) = g_{1}(\mathbf{\Theta}) + g_{2}(\mathbf{\Theta}) + g_{3}(\mathbf{\Theta}).
\label{IS-density-comp}
\end{equation}
Assuming that the point $\mathbf{\Theta}$ was drawn at iteration $i$,
$g_{1}$, $g_{2}$ and $g_{3}$ are the contributions to importance
weight for $\mathbf{\Theta}$ coming from iteration $i$, iterations
before $i$ and iterations after $i$ respectively. Thus, $g_{1}$ is calculated as follows:
\begin{equation} 
g_{1}(\mathbf{\Theta}) = \frac{q n_{i}}{N_{\mathrm{tot}} V_{{\rm tot},i}},
\label{IS-density-comp1}
\end{equation}
where $q$ is the number of ellipsoids at iteration $i$ in which point
$\mathbf{\Theta}$ lies, while $g_{2}$ is calculated as follows:
\begin{equation} 
g_{2}(\mathbf{\Theta}) = \frac{1}{N_{\mathrm{tot}}} \sum_{j=1}^{i-1} \frac{n_{j}}{V_{\rm tot},j},
\label{IS-density-comp2}
\end{equation}
where $V_{{\rm tot},j}$ is volume occupied by the union of ellipsoids at
iteration $j$ as given in Eq.~(\ref{IS-V}). Here we have assumed that
ellipsoids shrink at subsequent iterations and therefore points drawn
at iteration $i$ lie inside the ellipsoidal decompositions of previous
iterations as discussed earlier. Finally, $g_{3}$ is calculated as
follows:
\begin{equation} 
g_{3}(\mathbf{\Theta}) = \frac{1}{N_{\mathrm{tot}}} \sum_{j=i+1}^{n_{\mathrm{iter}}} \frac{n_{j} E_{j}(\mathbf{\Theta})}{V_{{\rm tot},j}}.
\label{IS-density-comp3}
\end{equation}

\subsection{Evidence error estimation}\label{INS:error}

As discussed by \citet{skilling04} (and by \citealt{feroz08,multinest}
for the specific case of {\sc MultiNest}) repeated summation
of the NS draws under random sampling of the associated $X_i$ (governed
by $t_i$; cf.\ Sec.~\ref{sec:multinest}) allows one to  estimate
the error on the NS evidence approximation from just a single run
(whereas many other MC integration techniques, such as thermodynamic
integration, require repeat runs to achieve this). Provided that the
parameter space has been explored with sufficient thoroughness (i.e.,
the $N_\mathrm{live}$ point set has evolved through all the
significant posterior modes), the
reliability of this evidence estimate was demonstrated in
\cite{feroz08}. Importantly, such a single run error estimate can also
be calculated for the INS scheme as described below.

Under ordinary (as opposed to pseudo-) importance sampling the
unbiased estimator for the asymptotic variance of the evidence
estimate here,
$\widehat{\mathrm{Var}}[\hat{\mathcal{Z}}]$, would be given as follows:
\begin{equation}
\widehat{\mathrm{Var}}[\hat{\mathcal{Z}}]
=
\frac{1}{N_{\mathrm{tot}}(N_{\mathrm{tot}}-1)}\sum_{k=1}^{N_{\mathrm{tot}}}\left[
  \frac{\mathcal{L}(\mathbf{\Theta}_{k})
    \pi(\mathbf{\Theta}_{k})}{g(\mathbf{\Theta}_{k})} - \mathcal{\hat{Z}} 
\right]^{2},
\label{eq:IS-var-Z}
\end{equation}
with $\mathcal{\hat{Z}}$ given by Eq.~(\ref{IS-Z}). 

With the draws from {\sc MultiNest} representing our
\textit{a posteriori} constructed $g(\mathbf{\Theta})$ not in fact an
independent, identically distributed sequence from this
pseudo-importance sampling function, the above uncertainty estimate is,
unfortunately, not strictly applicable here.  In particular, with the
placement of subsequent ellipses, $E_{j>i}$, dependent on the position
of the live particles drawn up to the present step, $i$, so too are
the subsequently drawn $\mathbf{\Theta}_{j>i}$.  However, when {\sc
MultiNest} is run in its default mode, such that we strongly govern
the maximum rate at which the volume of the successive $E_i$ can shrink we can
be confident that our sampling becomes ever more nearly independent
and that the dominant variance component is indeed given in
Eq.~(\ref{eq:IS-var-Z}).  Our reasoning behind this is explained in
detail in
Appendix B.  On the other hand, when {\sc MultiNest} is being run
in `constant efficiency mode' we recommended for the user to check
(via repeat simulation)
that the INS
evidence is stable (with respect to its error estimate) for reasonable variation in $N_{\rm
live}$ and/or $f$.

\section{Applications}\label{sec:applications}

In this section we apply the {\sc MultiNest} algorithm with INS
described above to three test problems to demonstrate that it indeed
calculates the Bayesian evidence much more accurately than vanilla
NS. These test examples are chosen to have features that
resemble those that can occur in real inference problems in astro- and
particle physics.

\subsection{Test problem 1: Gaussian shells likelihood}\label{sec:applications:shells}

%
\begin{figure*}
\psfrag{x}{$x$}
\psfrag{y}{$y$}
\psfrag{L}{$\mathcal{L}$}
\begin{center}
\subfigure[]{\includegraphics[width=0.47\columnwidth]{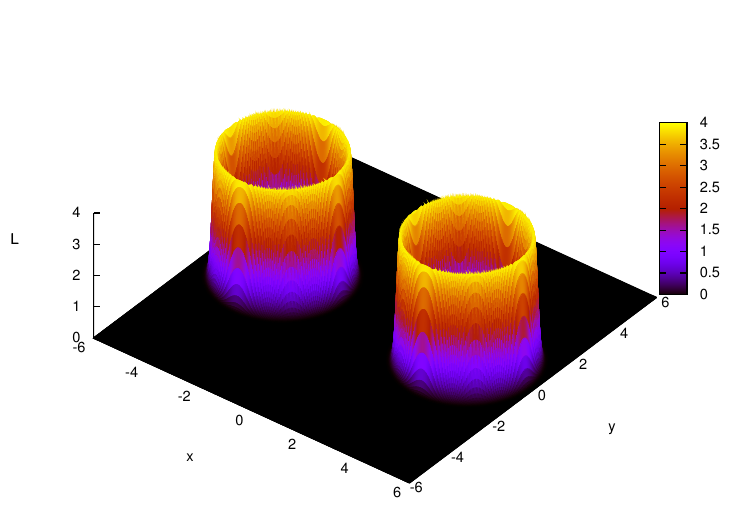}}\hspace{0.3cm}
\subfigure[]{\includegraphics[width=0.47\columnwidth]{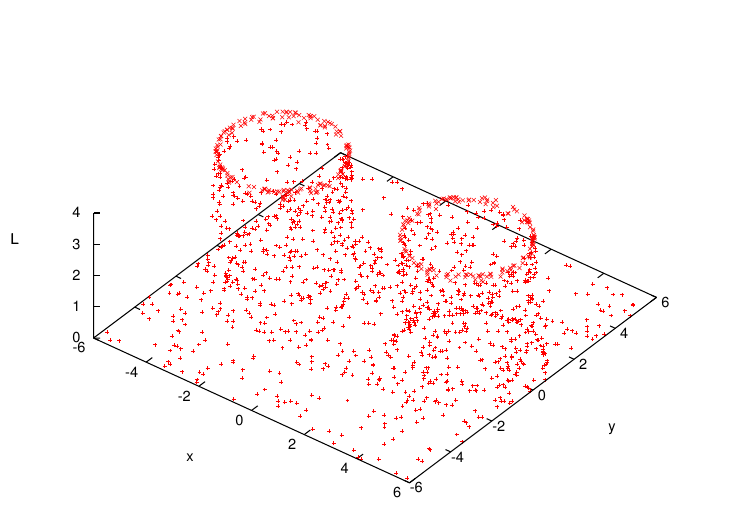}}
\caption{Test problem 1: (a) two-dimensional plot of the likelihood function defined in Eqs.~(\ref{eq:gshellL}) and (\ref{eq:gshell}); (b) dots denoting the points with the lowest likelihood at successive iterations of the {\sc MultiNest} algorithm.}
\label{fig:gshells}
\end{center}
\end{figure*}

In this section, we apply {\sc MultiNest} with and without INS to
sample from a posterior containing multiple modes with pronounced
(curving) degeneracies in relatively high dimensions. Our test problem
here is the same one used in \cite{bank, feroz08, multinest}. The
likelihood function of this problem is defined as,
\begin{equation}
\mathcal{L}(\mathbf{\theta})={\rm circ}(\mathbf{\theta};\mathbf{c}_1,r_1,w_1)+{\rm circ}(\mathbf{\theta};\mathbf{c}_2,r_2,w_2),
\label{eq:gshellL}
\end{equation}
where
\begin{equation}
{\rm circ}(\mathbf{\theta};\mathbf{c},r,w)=\frac{1}{\sqrt{2\pi
w^2}}\exp\left[-\frac{(\left|\mathbf{\theta}-\mathbf{c}\right|-r)^2}{2w^2}\right].
\label{eq:gshell}
\end{equation}

In two dimensions, this distribution represents two well separated rings, centred on the points $\mathbf{c}_1$ and $\mathbf{c}_2$ respectively, each of radius $r$ and with a Gaussian radial profile of width $w$ (see Fig.~\ref{fig:gshells}).

We investigate the above distribution up to a $50$-dimensional parameter space $\mathbf{\theta}$. In all cases, the centres of the two rings are separated by $7$ units in the parameter space, and we take $w_1=w_2=0.1$ and $r_1=r_2=2$. We make $r_1$ and $r_2$ equal, since in higher dimensions any slight difference between these two values would result in a vast difference between the volumes occupied by the rings and consequently the ring with the smaller $r$ value would occupy a vanishingly small fraction of the total probability volume, making its detection almost impossible. It should also be noted that setting $w=0.1$ means the rings have an extremely narrow Gaussian profile. We impose uniform priors $\mathcal{U}(-6, 6)$ on all the parameters. For the two-dimensional case, with the parameters described above, the likelihood is shown in Fig.~\ref{fig:gshells}.

\begin{table}
\begin{center}
\begin{tabular}{rrrrr}
\hline
$D$ & $N_{\rm live}$ & $f$ & $N_{\rm like}$ default & $N_{\rm like}$ ceff\\
\hline
 $2$& $300$ & $0.30$  &      $4,581$ &     $3,871$\\
 $5$& $300$ & $0.30$  &      $8,922$ &     $7,882$\\
$10$& $300$ & $0.05$ &     $73,342$ &    $76,255$\\
$20$& $300$ & $0.05$ &    $219,145$ &   $163,234$\\
$30$& $500$ & $0.05$ &    $604,906$ &   $548,501$\\
$50$& $500$ & $0.01$ & $10,531,223$ & $5,290,550$\\
\hline
\end{tabular}
\caption{Dimensionality ($D$) of problem, number of live points ($N_{\rm live}$), target efficiency ($f$) and the total number of likelihood evaluations ($N_{\rm like}$) in default and constant efficiency (ceff) modes of {\sc MultiNest} for test problem 1, discussed in Sec.~\ref{sec:applications:shells}.}
\label{tab:gshells_eff}
\end{center}
\end{table}

Table~\ref{tab:gshells_eff} lists the total number of live points
($N_{\rm live}$) and target efficiency ($f$) used and the total number
of likelihood evaluations ($N_{\rm like}$) performed by {\sc
MultiNest} in default and constant efficiency (ceff) modes. The volume
of the parameter space increases exponentially with the dimensionality
$D$, therefore we need to increase $N_{\rm live}$ and/or decrease $f$
with $D$, in order to get accurate estimates of $\log(\mathcal{Z})$. The
true and estimated values of $\log(\mathcal{Z})$ are listed in
Table~\ref{tab:gshells_evidence}.

\begin{table*}
\begin{center}
\begin{tabular}{rrrrrr}
\hline
     & Analytical & \multicolumn{2}{c}{{\sc MultiNest} without INS} & \multicolumn{2}{c}{{\sc MultiNest} with INS}\\
$D$  &          & default           & ceff              & default           & ceff\\
\hline
$2 $ & $  -1.75$ & $  -1.61 \pm 0.09$ & $  -1.71 \pm 0.09$ & $  -1.72 \pm 0.02$ & $  -1.69 \pm 0.02$\\
$5 $ & $  -5.67$ & $  -5.42 \pm 0.15$ & $  -5.78 \pm 0.15$ & $  -5.67 \pm 0.03$ & $  -5.87 \pm 0.03$\\
$10$ & $ -14.59$ & $ -14.55 \pm 0.23$ & $ -14.83 \pm 0.23$ & $ -14.60 \pm 0.03$ & $ -14.58 \pm 0.03$\\
$20$ & $ -36.09$ & $ -35.90 \pm 0.35$ & $ -35.99 \pm 0.35$ & $ -36.11 \pm 0.03$ & $ -36.06 \pm 0.03$\\
$30$ & $ -60.13$ & $ -59.72 \pm 0.35$ & $ -59.43 \pm 0.34$ & $ -60.09 \pm 0.02$ & $ -59.90 \pm 0.02$\\
$50$ & $-112.42$ & $-110.69 \pm 0.47$ & $-108.96 \pm 0.46$ & $-112.37 \pm 0.01$ & $-112.18 \pm 0.01$\\
\hline
\end{tabular}
\caption{The true and estimated $\log(\mathcal{Z})$ for test problem 1, discussed in Sec.~\ref{sec:applications:shells}, as a function of the dimensions $D$ of the parameter space, using {\sc MultiNest} with and without INS and in its default and constant efficiency modes.}
\label{tab:gshells_evidence}
\end{center}
\end{table*}

It can be seen from Table~\ref{tab:gshells_evidence} that
$\log(\hat{\mathcal{Z}})$ values obtained by {\sc MultiNest} with and
without INS and in both default and constant efficiency modes are
consistent with the true $\log(\mathcal{Z})$ for $D \le 20$, the only
exception being the $\log(\hat{\mathcal{Z}})$ from constant efficiency
mode with INS which is $\sim 6\sigma$ away from the analytical
$\log(\mathcal{Z})$. We attribute this to the heightened potential for
underestimation of the INS uncertainties in constant efficiency mode 
discussed in Sec.~\ref{sec:INS} and Appendix B. For $D \ge 30$ however, the
$\log(\hat{\mathcal{Z}})$ values obtained by {\sc MultiNest} without
INS start to become inaccurate, with constant efficiency mode again giving
more inaccurate results as expected. These inaccuracies are caused by
inadequate numbers of live points used to cover the region satisfying
the constraint $\mathcal{L}> \mathcal{L}_i$ at each iteration
$i$. However, with the same values for $N_{\rm live}$ and $f$, and
indeed with the same set of points, INS returns
$\log(\hat{\mathcal{Z}})$ values which are consistent with the true
$\log(\mathcal{Z})$ in default {\sc MultiNest} mode and off by at most
$\sim 0.2$ in the constant efficiency mode. The error estimate on
$\log(\hat{\mathcal{Z}})$ from INS in the constant efficiency mode
might indicate that $\log(\hat{\mathcal{Z}})$ has a large bias but as
discussed in Sec.~\ref{INS:error}, these error estimates are reliable
only when the importance sampling distribution is guaranteed to give
non-vanishing probabilities for all regions of parameter space where
posterior distribution has a non-vanishing probability as well. This
is much more difficult to accomplish in a $50$D parameter space. In
addition to this, the approximations we have made to calculate the
volume in the overlapped region of ellipsoids are expected to be less
accurate in higher dimensions. Therefore, it is very encouraging that
INS can obtain $\log(\mathcal{Z})$ to within $0.2$ units for a very
challenging $50$D problem with just $500$ live points. We should also
notice the number of likelihood evaluations in constant efficiency
mode starts to become significantly smaller than in the default mode
for $D \ge 20$.

\subsection{Test problem 2: egg-box likelihood}\label{sec:applications:eggbox}

%
\begin{figure*}
\psfrag{x}{$x$}
\psfrag{y}{$y$}
\psfrag{z}{$\log(\mathcal{L})$}
\psfrag{logL}{$\log(\mathcal{L})$}
\begin{center}
\subfigure[]{\includegraphics[width=0.47\columnwidth]{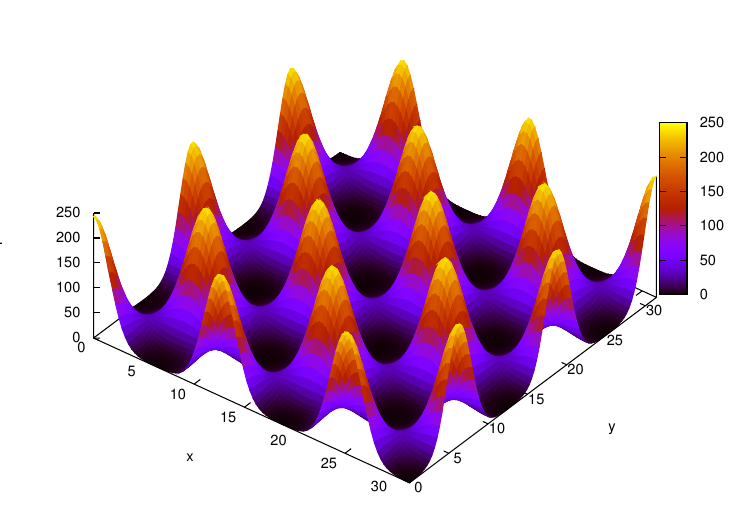}}\hspace{0.3cm}
\subfigure[]{\includegraphics[width=0.47\columnwidth]{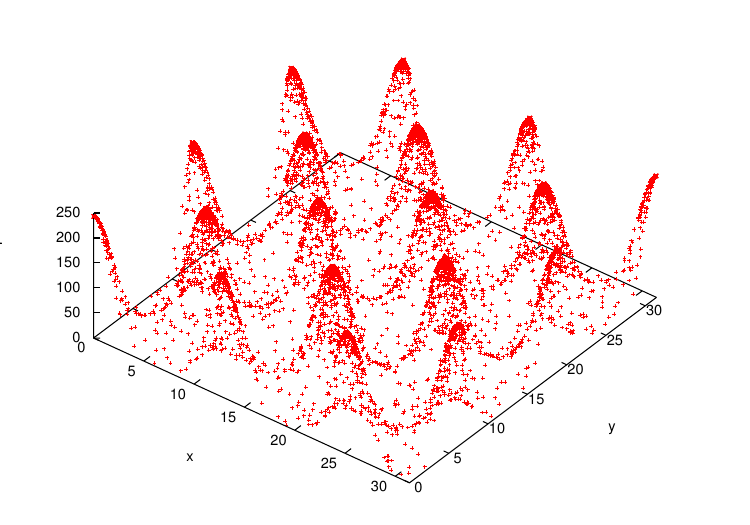}}
\caption{Test problem 2: (a) two-dimensional plot of the likelihood function defined in Eq.~\ref{eq:eggbox}; (b) dots denoting the points with the lowest likelihood at successive iterations of the {\sc MultiNest} algorithm.}
\label{fig:eggbox}
\end{center}
\end{figure*}

We now demonstrate the application of {\sc MultiNest} to a highly
multimodal two-dimensional problem, for which the likelihood resembles
an egg-box. The un-normalized likelihood is defined as:
\begin{equation}
\mathcal{L}(x, y) = \exp \displaystyle \left\{\left[2+\cos\left(\frac{x}{2}\right) \cos\left(\frac{y}{2}\right)\right]^5\right\},
\label{eq:eggbox}
\end{equation}
and we assume a uniform prior $\mathcal{U}(0,10\pi)$ for both $x$ and $y$.

A plot of the log-likelihood is shown in Fig.~\ref{fig:eggbox} and the
prior ranges are chosen such that some of the modes are truncated,
making it a challenging problem for identifying all the modes as well
as to calculate the evidence accurately. The true value of the
log-evidence is $\log Z = 235.856$, obtained by numerical integration
on a very fine grid, which is feasible for this simple two-dimensional
example.

It was shown in \citet{multinest} that {\sc MultiNest} can explore the
parameter space of this problem efficiently, and also calculate the
evidence accurately. Here we demonstrate the accuracy of the evidence
obtained with {\sc MultiNest} using the INS summation. For low-dimensional problems,
results obtained with the constant efficiency mode of {\sc MultiNest}
agree very well with the ones obtained with the default mode, we
therefore only discuss the default mode results in this section.

We use 1000 live points with target efficiency $f = 0.5$. The results
obtained with {\sc MultiNest} are illustrated in
Fig.~\ref{fig:eggbox}, in which the dots show the points with the
lowest likelihood at successive iterations of the nested sampling
process. {\sc MultiNest} required $\sim 20,000$ likelihood evaluations
and obtained $\log(\hat{\mathcal{Z}})=235.837 \pm 0.008$ ($235.848 \pm
0.078$) with (without) INS, which compares favourably with the true
value given above. In each case, the random number seed was the same,
so the points sampled by {\sc MultiNest} were identical with and
without INS. In order to check if the error estimates on
$\log(\hat{\mathcal{Z}})$ are accurate, we ran $10$ instances of {\sc
MultiNest} in both cases, each with a different seed and found the
mean and standard deviation of $\log(\hat{\mathcal{Z}})$ to be
$235.835 \pm 0.009$ ($235.839\pm 0.063$) with (without) INS.  In both
cases, the standard error agrees with the error estimate from just a
single run. There is, however, some indication of bias in the
$\log(\hat{\mathcal{Z}})$ value evaluated with INS, which lies $\sim
2\sigma$ away from the true value. This is most likely due to the
approximations used in calculating the volume in the overlapped region
of ellipsoids, as discussed in Sec.~\ref{sec:INS}. Nonetheless, the
absolute value of the bias is very low ($\sim 0.02$), particularly
compared with the accuracy ($\sim 0.5$) to which log-evidence values
are usually required in practical applications.

\subsection{Test problem 3: 16D Gaussian mixture model}\label{sec:applications:gmix}

%
\begin{figure*}
\psfrag{x}{$x$}
\psfrag{y}{$y$}
\psfrag{z}{$\log(\mathcal{L})$}
\begin{center}
\subfigure[]{\includegraphics[width=0.3\columnwidth]{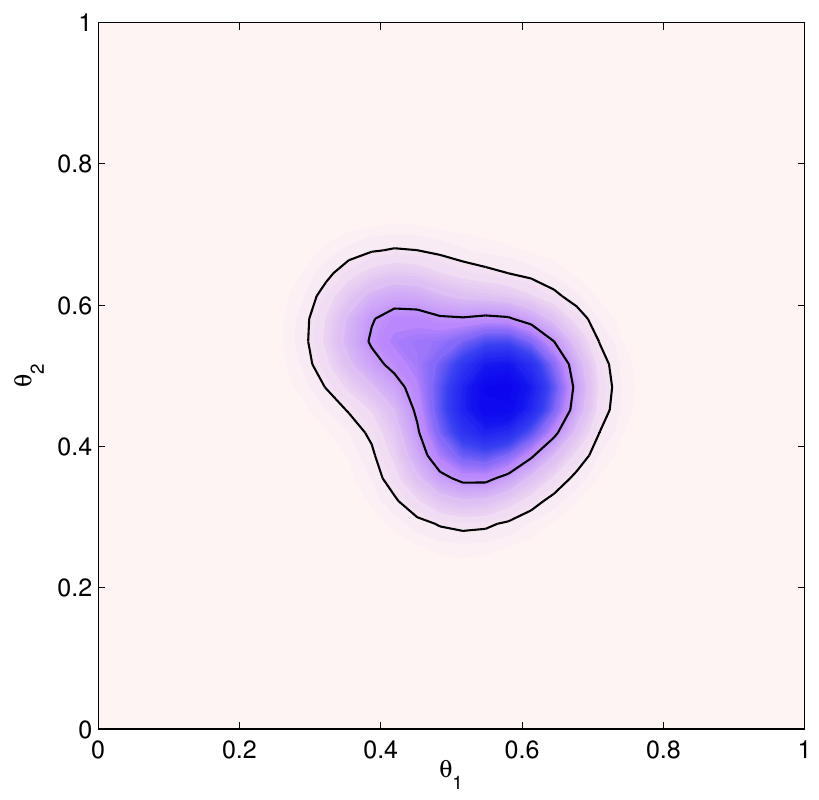}}\hspace{1cm}
\subfigure[]{\includegraphics[width=0.3\columnwidth]{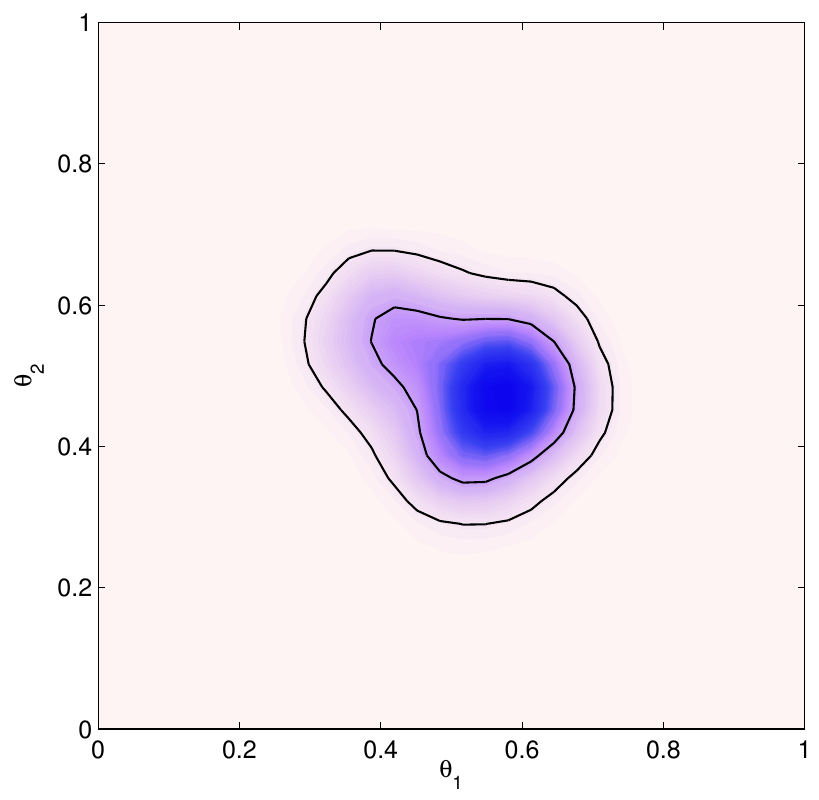}}
\caption{Test problem 3: Marginalized posterior distribution in the first 2 dimensions of the 16D Gaussian mixture model discussed in Sec.~\ref{sec:applications:gmix}. Panel (a) shows the analytical distribution while panel (b) shows the distribution obtained from {\sc MultiNest}. The contours represent the 68\% and 95\% Bayesian credible regions.}
\label{fig:weinberg}
\end{center}
\end{figure*}

Our next test problem is the same as test problem 4
in \citet{2013arXiv1301.3156W} which is a mixture model of four
randomly-oriented Gaussian distributions with their centers uniformly
selected from the hypercube $[0.5-2\sigma, 0.5+2\sigma]^{D}$ with $D$
being the dimensionality of the problem and the variance $\sigma^{2}$
of all four Gaussians is set to $0.003$. Weights of the Gaussians are
distributed according to a Dirichlet distribution with shape parameter
$\alpha = 1$. We impose uniform priors $\mathcal{U}(0, 1)$ on all the
parameters. The analytical posterior distribution for this problem,
marginalized in the first two dimensions is shown in
Fig.~\ref{fig:weinberg}(a).

The analytical value of $\log(\mathcal{Z})$ for this problem is $0$,
regardless of $D$. We set $D = 16$ and used $300$ live points with
target efficiency $f = 0.05$. The marginalized posterior distribution
in the first two dimensions, obtained with the default mode of {\sc
MultiNest} with INS is shown in Fig.~\ref{fig:weinberg}(b). The
posterior distribution obtained from the constant efficiency mode is
identical to the one obtained from the default and therefore we do not
show it. In the default mode {\sc MultiNest} performed $208,978$
likelihood evaluations and returned $\log(\hat{\mathcal{Z}})=-0.03 \pm
0.01$ ($0.39 \pm 0.27$) with (without) INS. In the constant efficiency
mode, $158,219$ likelihood evaluations were performed and
$\log(\hat{\mathcal{Z}}) = 0.21 \pm 0.01$ ($0.25 \pm 0.27$) with
(without) INS.

\subsection{Test problem 4: 20D Gaussian-LogGamma mixture model}\label{sec:applications:glgmix}

\begin{figure*}
\psfrag{x}{$\theta_1$}
\psfrag{y}{$\theta_2$}
\psfrag{z}{$\log(\mathcal{L})$}
\begin{center}
\subfigure[]{\includegraphics[width=0.35\columnwidth]{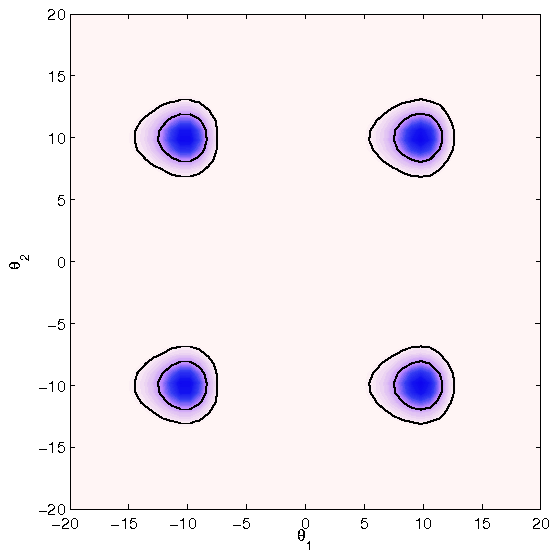}}\hspace{1cm}
\subfigure[]{\includegraphics[width=0.35\columnwidth]{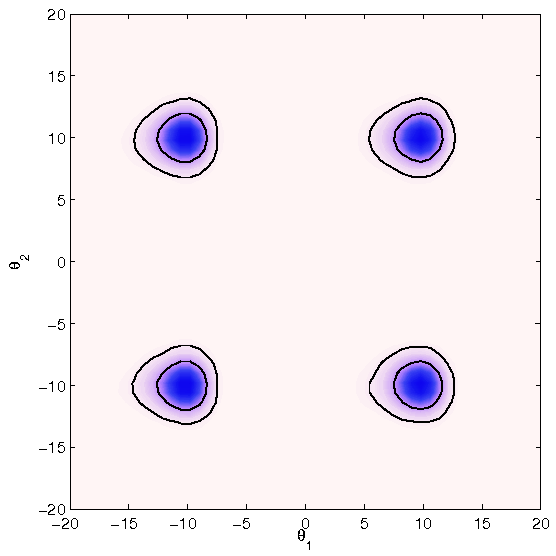}}
\caption{Test problem 4: Marginalized posterior distribution in the first 2 dimensions of the 20D Gaussian-LogGamma mixture model discussed in Sec.~\ref{sec:applications:glgmix}. Panel (a) shows the analytical distribution while panel (b) shows the distribution obtained from {\sc MultiNest}. The contours represent the 68\% and 95\% Bayesian credible regions.}
\label{fig:beaujean}
\end{center}
\end{figure*}

Our final test problem is the same as test problem 2 in 
\cite{Beaujean:arXiv1304.7808}, in which the likelihood 
is a mixture model consisting of four identical modes, each of which
is a product of an equal number of Gaussian and LogGamma 1D
distributions, centred at $\theta_1=\pm 10$, $\theta_2=\pm 10$,
$\theta_3=\theta_4=\cdots=\theta_D=0$ in the hypercube $\mathbf{\theta} \in
[-30,30]^D$, where $D$ is the (even) dimensionality of the parameter
space. Each Gaussian distribution has unit variance. The LogGamma
distribution is asymmetric and heavy-tailed; its scale and shape
parameters are both set to unity. We impose uniform priors ${\cal
U}(-30,30)$ on all the parameters. The analytical marginalised
posterior distribution in the subspace $(\theta_1,\theta_2)$ is shown
in Fig.~\ref{fig:beaujean}(a).

We set $D=20$, for which the analytical value of the log-evidence is
$\log(\mathcal{Z})=\log(60^{-20})=-81.887$. To be consistent with test
problem 3, which is of similar dimensionality, we again used 300 live
points with a target efficiency $f=0.05$ (note these values differ
from those used in \cite{Beaujean:arXiv1304.7808}, who set $N_{\rm
live}=1000$ and $f=0.3$ in the standard vanilla NS version of {\sc
MultiNest}). The marginalized posterior in the first two dimensions,
obtained in the default mode of {\sc MultiNest} with INS is shown in
Fig.~\ref{fig:beaujean}(b), and is identical to the corresponding
analytical distribution, recovering all four modes with very close to
equal weights. The posterior distribution obtained from the constant
efficiency mode is identical to the one obtained from the default and
therefore we do not show it. In the default mode {\sc MultiNest}
performed 2,786,538 likelihood evaluations and returned
$\log(\hat{\mathcal{Z}}) = -81.958 \pm 0.008$ ($-78.836 \pm 0.398$)
with (without) INS. In both cases, we see that, for this more
challenging problem containing multi-dimensional heavy-tailed
distributions, the log-evidence estimates are substantially biased,
with each being $\sim 8\sigma$ from the true value. Nonetheless, we
note that the estimate using INS is much more accurate than that
obtained with vanilla NS, and differs from the true value by only
$\sim 0.1$ units, which is much smaller than the accuracy required in most
practical applications. As one might expect, however, the log-evidence
estimates obtained in constant efficiency mode are somewhat poorer and
show a significant bias. In this mode, $297,513$ likelihood
evaluations were performed and $\log(\hat{\mathcal{Z}}) = -82.383 \pm
0.010$ ($-71.635 \pm 0.376$) with (without) INS.

\section{Summary and Discussion}\label{sec:discussion}

With the availability of vast amounts of high quality data,
statistical inference is increasingly playing an important role in
cosmology and astroparticle physics. MCMC techniques and more recently
algorithms based on nested sampling have been employed successfully in
a variety of different areas. The {\sc MultiNest} algorithm in
particular has received much attention in astrophysics, cosmology and
particle physics owing to its ability to efficiently explore
challenging multi-modal distributions as well as to calculate the
Bayesian evidence.

In this paper we have discussed further development of the {\sc
MultiNest} algorithm, based on the implementation of the INS scheme
recently proposed by \cite{cam13}. INS requires no
change in the way {\sc MultiNest} explores the parameter space, but
can calculate the Bayesian evidence at up to an order-of-magnitude
higher accuracy than vanilla nested sampling. Moreover, INS also
provides a means to obtain reasonably accurate evidence estimates from
the constant efficiency mode of {\sc MultiNest}. This is particularly
important, as the constant efficiency mode enables {\sc MultiNest} to
explore higher-dimensional spaces (up to $\sim 50D$) much more
efficiently than the default mode. Higher evidence accuracy from INS
could potentially allow users to use fewer live points $N_{\rm live}$
or higher target efficiency $f$ to achieve the same level of accuracy
as vanilla nested sampling, and therefore to speed-up the analysis by
several factors. We recommend that users should always check that
their posterior distributions are stable with reasonable variation of
$N_{\rm live}$ and $f$. A slight drawback of INS is increased memory
requirements. As the importance sampling distributions given in
Eqs.~\ref{IS-density-comp2} and \ref{IS-density-comp3} change for
every point at each iteration, all the points need to be saved in
memory. However, with $N_{\rm live} \le 1000$ the increased memory
requirements should be manageable on most modern computers.

Finally, we give some recommendations for setting the number of live points $N_{\rm live}$ and target efficiency $f$, which determine the accuracy and computational cost of running the {\sc MultiNest} algorithm, with or without INS. Generally, the larger the $N_{\rm live}$ and lower the $f$, the more accurate are the posteriors and evidence values but the higher the computational cost. For multi-modal problems, $N_{\rm live}$ is particularly important as it determines the effective sampling resolution. If it is too small, certain modes, in particular the ones occupying a very small prior mass, can be missed. Experience has shown that the accuracy of evidence is more sensitive to $f$ than $N_{\rm live}$. In general, for problems where accuracy of evidence is paramount, we suggest $f$ to be no larger than $0.3$ in the `default' mode. In `constant efficiency mode', we suggest $f$ to be no larger than $0.1$ in all cases. Generally, a value of $N_{\rm live}$ in lower hundreds is sufficient. For very low dimensional problems $N_{\rm live}$ can even be in tens. However, for highly multi-modal problems, one may need to set $N_{\rm live}$ to be in a few thousands. It is always advisable to increase $N_{\rm live}$ and reduce $f$ to check if the posteriors and evidence values are stable as function of $N_{\rm live}$ and $f$.

\section*{Acknowledgements}\label{sec:ackn}
This work was performed on COSMOS VIII, an SGI Altix UV1000
supercomputer, funded by SGI/Intel, HEFCE and PPARC, and the authors
thank Andrey Kaliazin for assistance. The work also utilized the
Darwin Supercomputer of the University of Cambridge High Performance
Computing Service (\texttt{http://www.hpc.cam.ac.uk/}), provided by
Dell Inc. using Strategic Research Infrastructure Funding from the
Higher Education Funding Council for England. FF is supported by a
Research Fellowship from the Leverhulme and Newton Trusts. EC is
supported by an Australian Research Council (ARC) Discovery Grant.
ANP is supported by an ARC Professorial Fellowship.

\begin{appendices}

\section{Relation of INS to Existing Monte Carlo Schemes}\label{sec:appendixa}

We review here the heritage of INS amongst the wider family of
pseudo-importance sampling, NS, and adaptive Monte Carlo algorithms,
for which limited convergence proofs have yet been achieved.


As described in \citet{cam13} the initial idea for INS arose from the
study of recursive marginal likelihood estimators, as characterised by
Reverse Logistic Regression
(RLR; \citealt{gey94}; \citealt{che97}; \citealt{kon03}) and the
Density of States (DoS; \citealt{hab12}; \citealt{tan12}).  In these
(\textit{equivalent}; cf.\ \citealt{cam13}) schemes, the marginal
likelihood is sought by pooling (or `losing the labels' on) a series
of draws from a pre-specified set of largely unnormalised importance
sampling densities, bridging (at least crudely) the prior and
posterior; after this a maximum-likelihood-based estimator is used to
infer the relative normalisation of each bridging density in light of
the `missing' label information.  As emphasised by \citet{kon03}, these
recursive algorithms may, in turn, be seen as deriving from the
`biased sampling' results of
\citet{var85} and collaborators (e.g.\ \citealt{gil98}),
who give consistency and Central Limit Theorem proofs for this
deterministic (i.e., non-adaptive), \textit{pseudo-importance
sampling} procedure under simple connectedness/non-separability
conditions for the supports of the bridging sequence.

Developed in parallel to the recursive estimators described above, the
Deterministic Multiple Mixture Sampling scheme
(DMMS; \citealt{vea95}; \citealt{owe00}) applies much the same
strategy, but for a given sequence of strictly \textit{normalised}
importance sampling proposal densities; hence, the motivation for
`losing the labels' here becomes simply the reduction of variance in
the resulting estimator with respect to that achievable by allocating
the same number of draws to ordinary importance sampling from each
proposal separately.  [We will discuss further the limiting variance
of a simple `losing the labels' estimator, as relevant to INS, in 
Appendix B.]  At the expense of introducing an \textit{intractable} (but
asymptotically-diminishing) \textit{bias},
\citet{cor12} have recently constructed a yet more efficient extension 
called Adaptive Multiple Importance Sampling (AMIS) in which the
sequence of importance sampling proposal densities is refined
adaptively towards the target at
runtime.  In particular, each proposal density after the first is
chosen (from a given parametric family) in a manner dependent on the
weighted empirical distribution of draws from all previous densities.
As suggested by their given
numerical examples this approach
appears superior to other adaptive importance sampling schemes (e.g.\
the cross-entropy method; cf.\ \citealt{rub04}) in
which the past draws from sub-optimal proposals are ultimately
discarded.

In our opinion, despite its genesis in the study of RLR/DoS, INS may perhaps most accurately be viewed as a `descendent' of this AMIS
methodology; the key difference being that INS builds an efficient mixture
proposal density for marginal likelihood estimation via the NS pathway
(Sec.\ 3), whereas AMIS aims to iterate towards such a form
chosen from within a pre-specified family of proposal
densities.  In other words, in INS the proposal densities represented by our
sequence of 
ellipsoidal decompositions should share (by design) a near-equal
`importance' in the final mixture, while in AMIS those draws from the
earlier proposals are expected to become increasingly insignificant as
the later proposals achieve refinement towards their target.

As acknowledged by \citet{cor12}, the inherent dependence structure of
the pseudo-importance weighted terms entering the final AMIS
summation---owing entirely in this approach to the dependence
structure of the corresponding sequence of proposal
densities---renders intractable the demonstration of a
\textit{general} consistency for the algorithm.  Indeed
even the elegant and detailed solution presented by \citet{mar12}
is reliant on key modifications to the basic procedure, including that the number of points drawn from each
successive density grows significantly at each iteration (incompatible
with INS; and seemingly at odds too with Cornuet et al.'s original
recommendation of heavy sampling from the first proposal), as well as numerous
assumptions regarding the nature of the target and proposal families.
In light of this historical background we will therefore give particular
attention in the following Appendix to the variance reduction benefits of the
theoretically-problematic 
`losing the labels' strategy as employed in INS, before sketching a rough proof of consistency thereafter (albeit under some strong assumptions on the
asymptotic behaviour of the EM plus $k$-means algorithm employed for ellipsoidal
decomposition with respect to the target density; which may be difficult, if not impossible,
to establish in practice).

Finally, before proceeding it is worth mentioning briefly the heritage
of INS with respect to vanilla NS \citep{skilling04, ski06}. As described in
Sections 3 and 4, the original {\sc MultiNest} code
(\citealt{feroz08}; \citealt{multinest}) was designed for estimation of $\mathcal{Z}$
via the NS pathway \citep{ski06} with the challenge of
constrained-likelihood sampling tackled via rejection sampling within
a series of ellipsoidal decompositions bounding the evolving live
point set.  In contrast to AMIS (and INS) the convergence properties
of the simple NS algorithm are well understood; in
particular, \citet{cho10} have derived a robust CLT
for nested sampling, and both \citet{ski06} and \citet{kee11} give
insightful discussions. Despite the value of this ready availability
of a CLT for vanilla NS, our experience (cf.\ Sec.\ 5 of the main
text) is that by harnessing the information content of the
otherwise-discarded draws from {\sc MultiNest}'s ellipsoidal rejection
sampling the INS summation does ultimately yield in practice a
substantially more efficient approximation to the desired marginal
likelihood, the reliability of which is moreover adequately estimable
via the prescription given subsequently.

\section{Convergence Behaviour of INS}\label{sec:appendixb}

In this Appendix we discuss various issues relating to the
\textit{asymptotic convergence} of the INS marginal likelihood 
estimator (\ref{IS-Z}), which we denote here by
$\hat{\mathcal{Z}}^\mathrm{INS}$, towards the true marginal likelihood, $\mathcal{Z}$, as
the sample size (controlled by the size of the live point set,
$N_\mathrm{live}$) approaches infinity. We begin by considering the
intriguing role of pseudo-importance sampling for variance reduction
within certain such schemes; this step, ironically, is itself
primarily responsible for the intractable bias of the complete
algorithm.  With this background in mind we can at last outline a
heuristic argument for the consistency of INS and consider a break down of its variance into distinct terms of transparent origin.

To be precise, we will investigate here the asymptotic convergence
behaviour of the INS estimator with ellipsoidal decompositions almost
exactly as implemented in
{\sc MultiNest}, a detailed description of which is given in the
main text (Sections 3 \& 4).  For reference, we take 
\begin{eqnarray}
\hat{\mathcal{Z}}^\mathrm{INS}\ 
&\equiv&
\frac{1}{N_\mathrm{tot}} \sum_{i=1}^{c \times N_\mathrm{live}}
\sum_{k=1}^{n_i}
\frac{\mathcal{L}(\bm{\Theta}_k^{(i)})\pi(\bm{\Theta}_k^{(i)})}{g(\bm{\Theta}_k^{(i)})}, \label{INSdef} \nonumber\\
&=&\ \frac{1}{N_\mathrm{tot}}\sum_{k=1}^{N_\mathrm{tot}}\frac{\mathcal{L}(\bm{\Theta}_k)\pi(\bm{\Theta}_k)}{g(\bm{\Theta}_k)}, \label{INSdef2}
\end{eqnarray}
with $N_\mathrm{tot}\equiv\sum_{i=1}^{c \times N_\mathrm{live}} n_i$,
 $g(\bm{\Theta}) \equiv \frac{1}{N_\mathrm{tot}} \sum_{i=1}^{ c \times
 N_\mathrm{live}} \frac{n_i E_i(\bm{\Theta})}{V_{{\rm tot},i}}$, and $c \times
 N_\mathrm{live}$ a fixed stopping proxy for the total number of
 ellipsoidal decompositions required to meet our actual flexible
 stopping criterion (cf.\ Sec.\ 3.2). Each collection of
 $\bm{\Theta}_{k=1,\ldots,n_i}^{(i)}$ here is assumed drawn uniformly
 from the corresponding ellipsoidal decomposition (of the live
 particle set), $E_i(\cdot)$, with volume, $V_{{\rm tot},i}$, until the discovery
 of a single point, say $\bm{\Theta}_{n_i}^{(i)}$, with $
\mathcal{L}(\bm{\Theta}_{n_i}^{(i)}) > \mathcal{L}_{i-1}$.  This new
constrained-likelihood point serves, of course, as the replacement to
the $\mathcal{L}_{i-1}$ member of the NS live point set against which
the next, $E_{i+1}(\cdot)$, decomposition is then defined.  

The equality in (\ref{INSdef2}) highlights the fact that having pooled
our draws from each $E_i(\cdot)$ into the pseudo-importance sampling
function, $g(\cdot)$, we may proceed to `lose the labels', ${}^{(i)}$,
on these as in, e.g., Reverse Logistic Regression or ``biased
sampling''.  Note also that we suppose $E_1(\cdot)$ is fixed to the
support of the 
prior itself (to ensure that the support of $\mathcal{L}(\bm{\Theta})\pi(\bm{\Theta})$ is contained
within that of $g(\bm{\Theta})$), and that we must sample an initial
collection of $N_\mathrm{live}$ live points from the prior as well
to populate the original live particle set in advance of our first
constrained-likelihood exploration.  

Finally, we neglect in the ensuing analysis any uncertainty in our
$V_{{\rm tot},i}$ since, although these are in fact estimated also via
(simple MC) simulation, without the need for likelihood function calls
in this endeavour we consider the cost of arbitrarily improving their
precision effectively negligible.

\subsection{Motivation for `losing the labels' on a normalised pseudo-importance sampling mixture}

The effectiveness of the so-called `losing the labels' strategy for marginal
likelihood estimation via the recursive pathway can be
easily appreciated for the typical RLR/DoS case of multiple unnormalised bridging
densities, since by allowing for, e.g., the use of tempered Monte Carlo
sampling we immediately alleviate to a large extent the burdens of
importance sampling 
 proposal design (cf.\ \citealt{hes95}).  However, its utility in cases of strictly normalised mixtures of
 proposal densities as encountered in DMMS and INS is perhaps surprising.
\citet{owe00} give a proof that, under the DMMS scheme, the asymptotic
 variance of the final estimator will not be very much worse than that
 achievable under ordinary importance sampling from the optimal
 distribution alone.  However, as the INS sequence
 of ellipsoids is not designed to contain a single optimal proposal,
 but rather to function `optimally' as an ensemble we focus here
 on demonstrating the strict ordering (from largest to smallest) of the asymptotic variance for
 \textsc{(i)} ordinary importance sampling under each mixture
 component separately, \textsc{(ii)} ordinary importance sampling
 under the true mixture density itself, and \textsc{(iii)}
 \textit{pseudo}-importance sampling from the mixture density (i.e., `losing
 the labels').

Consider a grossly simplified version of INS in which, at the $n$th
iteration (it is more convenient here to use $n$ rather than $i$ as
the iteration counter), a single random point is drawn independently
from each of $n$ labelled densities, $h_{k,n}(\cdot)$
$(k=1,2,\ldots,n)$, with identical supports matching those of the
target, $f(\cdot)$.  We denote the resulting set of $n$ samples by
$\bm{\Theta}^{(n)}_{k=1,2,\ldots,n}$. The three key simplifications
here are: \textsc{(i)} that the draws are independent, when
in {\sc MultiNest} they are inherently dependent; \textsc{(ii)}
that the supports of the $h_{k,n}(\cdot)$ match, when in fact the
ellipsoidal decompositions, $E_n(\cdot)$, of {\sc MultiNest} have
generally nested supports (though one \textit{could} modify them
appropriately in the manner of defensive sampling; \citealt{hes95});
and \textsc{(iii)} that a single point is drawn from each labelled
density, when in fact the sampling from each $E_n(\bm{\Theta})$ under
{\sc MultiNest} follows a negative binomial distribution for one
$E_n(\bm{\Theta})\cap \{\mathcal{L}(\bm{\Theta})>\mathcal{L}_{n-1}\}$
`success'.  Suppose also now that the unbiased estimator,
$\hat{\mathcal{Z}}_k^{(n)}=f(\bm{\Theta}_k^{(n)})/h_{k,n}(\bm{\Theta}_k^{(n)})$,
for the normalizing constant belonging to $f(\cdot)$, namely $\mathcal{Z}=\int
f(\bm{\Theta}) d\bm{\Theta}$, in such single draw importance sampling
from each of the specified $h_{k,n}(\cdot)$ has finite (but non-zero)
variance (cf.\ \citealt{hes95}), i.e.,
\begin{equation}
\sigma_{k,n}^2 = \int \frac{f(\bm{\Theta})^2}{h_{k,n}(\bm{\Theta})}
 d\bm{\Theta} -\mathcal{Z}^2,\ \qquad 0 < \sigma_{k,n}^2 < \infty,
\end{equation}
\textit{and} that
 together our $\hat{\mathcal{Z}}_k^{(n)}$ satisfy Lindeberg's
 condition such that the CLT holds for this triangular array of random
 variables (cf.\ \citealt{bil95}).

Now, if we would decide to \textbf{keep the labels}, $k$, on
our independent draws from the sequence of $h_{k,n}(\cdot)$ then
supposing no prior knowledge of any $\sigma_{k,n}^2$ (i.e., no prior
knowledge of how close each proposal density might be to our target)
the most sensible option might be to take as a `best guess' for $\mathcal{Z}$
the (unweighted) sample mean of our individual
$\hat{\mathcal{Z}}_k^{(n)}=f(\bm{\Theta}^{(n)}_k)/h_{k,n}(\bm{\Theta}^{(n)}_k)$,
that is, 
\begin{equation}
\hat{\mathcal{Z}}_\mathrm{labelled} = \frac{1}{n} \sum_{k=1}^n \frac{f(\bm{\Theta}_k^{(n)})}{h_{k,n}(\bm{\Theta}_k^{(n)})}.
\end{equation}
With a common mean and finite variances for each, this sum over a triangular array
converges (in distribution) to a univariate Normal with mean,
$\mathcal{Z}$, and variance, 
\begin{equation}
\sigma^2_\mathrm{labelled} = \frac{s_n^2}{n} = \frac{1}{n} \sum_{k=1}^n \left[\int f(\bm{\Theta})^2/h_{k,n}(\bm{\Theta}) d\bm{\Theta} - \mathcal{Z}^2\right],
\end{equation}
here we use the abbreviation,  $s_n^2 = \sum_{k=1}^n \sigma_{k,n}^2$.

On the other hand, if we would instead decide to \textbf{lose the
  labels} on our independent draws we might then follow Vardi's
method and imagine each $\bm{\Theta}_k^{(n)}$ to have come from the mixture
distribution, $g(\bm{\Theta}) = \frac{1}{n} \sum_{j=1}^n h_{j,n}(\bm{\Theta})$,
for which the
alternative estimator,
\begin{equation}
\hat{\mathcal{Z}}_\mathrm{unlabelled} = \frac{1}{n} \sum_{k=1}^n
\frac{f(\bm{\Theta}_k^{(n)})}{g(\bm{\Theta}_k^{(n)})},
\end{equation}
 may be derived. To see that the
$\hat{\mathcal{Z}}_\mathrm{unlabelled}$ estimator so defined is in fact unbiased
we let $\hat{\mathcal{Z}}_k^{(n)\prime} =
f(\bm{\Theta}^{(n)})/h_{k,n}(\bm{\Theta}^{(n)})$ for
$\bm{\Theta}^{(n)} \sim h_{k,n}(\cdot)$ and observe that
\begin{eqnarray}
\mathrm{E}[\hat{\mathcal{Z}}_\mathrm{unlabelled}] 
&=& \frac{1}{n} \sum_{k=1}^n \mathrm{E}[\hat{\mathcal{Z}}_k^{(n)\prime}], \nonumber \\
&=& \frac{1}{n} \sum_{k=1}^n \int \frac{f(\bm{\Theta})h_{k,n}(\bm{\Theta})}{g(\bm{\Theta})} d\bm{\Theta} \nonumber, \\
&=& \int \frac{f(\bm{\Theta})}{g(\bm{\Theta})}\left[\frac{1}{n}\sum_{k=1}^n h_{k,n}(\bm{\Theta})\right] d\bm{\Theta} = \mathcal{Z}.
\end{eqnarray}
For $n$ iid samples drawn
faithfully from the mixture density, $g(\cdot)$, we would expect (via the CLT)
that the estimator
$\hat{\mathcal{Z}}_\mathrm{unlabelled}$ will converge (in distribution) once
again to a univariate Normal
with mean, $\mathcal{Z}$, but with alternative variance,
$\sigma_\mathrm{unlabelled}^{2\ [g(\cdot),\mathrm{true}]} = \frac{1}{n} \int
f(\bm{\Theta})^2/g(\bm{\Theta})  d\bm{\Theta} - \mathcal{Z}^2$.  However, for the
pseudo-importance sampling from $g(\cdot)$ described above, in which
we instead pool an explicit 
sample from each of its separate mixture components, the asymptotic
variance of $\hat{\mathcal{Z}}_\mathrm{unlabelled}$ is significantly smaller again.  In
particular, 
\begin{eqnarray}
\sigma_\mathrm{unlabelled}^{2\ [g(\cdot),\mathrm{pseudo}]}
&=&\sum_{k=1}^n \mathrm{Var}[\hat{\mathcal{Z}}_k^{(n)\prime}/n], \nonumber \\
&=& \frac{1}{n^2} \sum_{k=1}^n \{\mathrm{E}[(\hat{\mathcal{Z}}_k^{(n)\prime})^2] - \mathrm{E}[\hat{\mathcal{Z}}_k^{(n)\prime}]^2  \}, \nonumber \\
&=& \frac{1}{n^2} \sum_{k=1}^n \int\frac{f(\bm{\Theta})^2}{g(\bm{\Theta})^2} h_{k,n}(\bm{\Theta}) d\bm{\Theta} - \frac{1}{n^2} \sum_{k=1}^n \mathrm{E}[\hat{\mathcal{Z}}_k^{(n)\prime}]^2, \nonumber \\
&=& \frac{1}{n} \int \frac{f(\bm{\Theta})^2}{g(\bm{\Theta})} d\bm{\Theta} - \frac{1}{n^2} \sum_{i=1}^n \{(\mathrm{E}[\hat{\mathcal{Z}}_k^{(n)\prime}] - \mathcal{Z}) + \mathcal{Z}\}^2, \nonumber \\
&=& \sigma_\mathrm{unlabelled}^{2\ [g(\cdot),\mathrm{true}]} - \frac{1}{n^2} \sum_{i=1}^n \{\mathrm{E}[\hat{\mathcal{Z}}_k^{(n)\prime}] - \mathcal{Z}\}^2, \nonumber \\
&\le&\sigma_\mathrm{unlabelled}^{2\ [g(\cdot),\mathrm{true}]},
\end{eqnarray}
with equality achieved only in the trivial case that all mixture
components are identical.  The variance reduction here in the
pseudo-importance sampling framework relative to the true importance
sampling case derives of course from the effective replacement of
multinomial sampling of the mixture components by fixed sampling from
their expected proportions.

Comparing now the asymptotic variances of our labelled and unlabelled
estimators we can see that the latter is (perhaps
surprisingly) \textit{always} smaller than the former, i.e.,
\begin{equation}
\sigma_\mathrm{unlabelled}^{2\ [g(\cdot),\mathrm{true}]} - \sigma_\mathrm{labelled}^{2} = {\displaystyle\frac{1}{n}\int f(\bm{\Theta})^2 \left[\sum_{k=1}^n \frac{1}{h_{k,n}(\bm{\Theta})} - \frac{n}{(\sum_{k=1}^n h_{k,n}(\bm{\Theta}))} \right] d\bm{\Theta} < 0},
\end{equation}
(recalling that all densities here are strictly positive, of course);
thus, we observe the ordering, \[\sigma_\mathrm{unlabelled}^{2\ [g(\cdot),\mathrm{pseudo}]} <
\sigma_\mathrm{unlabelled}^{2\ [g(\cdot),\mathrm{true}]} < \sigma_\mathrm{labelled}^{2}.\]  This is, as has been
remarked in the past, the paradox of the `losing the labels' idea; that \textit{by
throwing away information about our sampling process we appear to \textbf{gain}
information about our target}!  In fact, however, all we are really
doing by choosing to estimate $\mathcal{Z}$ with $\hat{\mathcal{Z}}_\mathrm{unlabelled}$
rather than $\hat{\mathcal{Z}}_\mathrm{labelled}$ is to use the information we
have extracted from $f(\cdot)$ in a more efficient manner, as
understood (from
the above error analysis) \textit{a priori} to our actual importance
sampling.  The strict ordering shown here explains
why we have selected a pseudo-importance sampling strategy for
combining the ellipsoidal draws in {\sc MultiNest} as opposed to,
e.g., modifying our sampling from the $E_n(\cdot)$ to match
(defensively) the support of $\pi(\bm{\Theta})$ and compiling
estimators $\hat{\mathcal{Z}}_k$ separately---though the latter would simplify our
convergence analysis the former should be (asymptotically) much more efficient.

\subsection{Consistency of INS} 

To establish a heuristic argument for
 consistency of the INS scheme we must first consider the nature of
the limiting distribution for the sequence of ellipsoidal
decompositions, $\{E_i(\cdot)\}$, as $N_\mathrm{live} \rightarrow
\infty$.  To do so we introduce the following strong assumption: that
for the constrained-likelihood contour corresponding to each $X_i \in
[0,1]$ there exists a unique, limiting ellipsoidal decomposition,
$E^\ast_{X_i}(\cdot)$, to which the {\sc MultiNest} algorithm's
$E_i(\cdot)$ will converge (if not stopped early)
\textit{almost surely} for all $N_\mathrm{live}$ for which $X_i
=\exp(-i/N_\mathrm{live})$ for some $i \in \{1,2,\ldots,c\times
N_\mathrm{live}\}$.  In particular, we suppose that both the design of
the EM plus $k$-means code for constructing our $E_i(\cdot)$ and the
nature of the likelihood function, $\mathcal{L}(\bm{\Theta})$, are
such for any given $\epsilon >0$ there is an $N_\mathrm{live}$ large
enough that thereafter
\[\sup_{i} \left(
\frac{\mathrm{P}_{\pi(\bm{\Theta})}\{\bm{\Theta} \in E_{i}(\cdot) \cap
E^\ast_{X_i=\exp(-i/N_\mathrm{live})}(\cdot)\}}{\mathrm{P}_{\pi(\bm{\Theta})}\{\bm{\Theta} \in E_{i}(\cdot) \cup
E^\ast_{X_i=\exp(-i/N_\mathrm{live})}(\cdot)\}}\right) > 1- \epsilon.\]
Another supposition we make is that the limiting family of ellipsoidal
decompositions, $\{E^\ast_{X_i}(\cdot)\}$, is at least left or right 
`continuous' in the
same sense at every point of its rational baseline; i.e., for each
$X_i$ and any $\epsilon>0$ there exists a $\delta>0$ such that
$X_i-X_j < \delta$ and/or $X_j-X_i < \delta$ implies 
\[\frac{\mathrm{P}_{\pi(\bm{\Theta})}\{\bm{\Theta} \in E^\ast_{X_i}(\cdot) \cap
E^\ast_{X_j}(\cdot)\}}{\mathrm{P}_{\pi(\bm{\Theta})}\{\bm{\Theta} \in E^\ast_{X_i}(\cdot) \cup
E^\ast_{X_j}(\cdot)\}} > 1- \epsilon.\]

Various conditions for almost sure convergence of 
EM \citep{wu83} and $k$-means \citep{pol81} algorithms have been
demonstrated in the past, but we have an intractable dependence
structure operating on the $E_k(\cdot)$ for INS and it is not at all
obvious how to clearly formulate such conditions here.  The complexity
of this task can perhaps most easily be appreciated by considering the
limited availability of results for the convergence in volume of
random convex hulls from uniform sampling within regular polygons in
high-dimensions (e.g.\ \citealt{sch80}; \citealt{sch08}).  On the
other hand, we may suspect that the necessary conditions for the above
are similar to those required in any case for almost sure `coverage'
of each constrained-likelihood surface by its corresponding
ellipsoidal decomposition; the latter being an often ignored
assumption of rejection NS.  That is, even for a generous dilation of
the simple proposal ellipsoids, as suggested by
\citet{muk06}, one can easily identify some family of
(typically non-convex)
$\mathcal{L}(\bm{\Theta})$ for which the given dilation factor will be
demonstrably insufficient; though whether such a `pathological'
$\mathcal{L}(\bm{\Theta})$ would be likely to arise in standard statistical
settings is perhaps another matter entirely!

The necessity of these assumptions, and in particular our second
regarding the `continuity' of the limiting
$\{E^\ast_{X_i}(\cdot)\}$, is two-fold: to ensure that a limiting
distribution exists (this echoes the requirement for there to exist an
optimal proposal in the equivalent AMIS analysis
of \citealt{mar12}), \textit{and} to ensure that its form is such as
to render irrelevant the inevitable stochastic variation and bias in
our negative binomial sampling of $n_i$ points from $E_i(\cdot)$.
Important to acknowledge is that not only is the number of points
drawn from each ellipsoidal decomposition, $n_i$, a random variable,
but the collection of $n_i-1$ draws in
$E_i(\cdot)\cap\{\mathcal{L}(\bm{\Theta})<\mathcal{L}_{i-1}\}$ plus
one in $E_i(\cdot)\cap\mathcal{L}(\bm{\Theta})>\mathcal{L}_{i-1}$ from
a single realization cannot strictly be considered a uniform draw from
$E_i(\cdot)$, though we treat it as such in our summation for
$g(\bm{\Theta})$.  Indeed the expected proportion of these draws
represented by the single desired
$\mathcal{L}(\bm{\Theta})>\mathcal{L}_{i-1}$ point,
namely $\mathrm{E}[\frac{1}{n_i}] = \frac{-p_i\log p_i}{1-p_i}$, does not
even match its fraction of $\pi(\bm{\Theta})$ by `volume', here $p_i$.
Our argument must therefore be that (asymptotically) with more and
more near-identical ellipsoids converging in the vicinity of each
$E^\ast_i(\cdot)$ as $N_\mathrm{live} \rightarrow \infty$
the \textit{pool} of all such biased draws from our
constrained-likelihood sampling within each of these nearby,
near-identical ellipsoids ultimately approximates an unbiased
sample, \textit{and} that the mean number of draws from these will
approach its long-run average, say $n^\ast_i$.
 
With such convergence towards a limiting distribution,
$F^\ast(\bm{\Theta})$, defined by the set of pairings,
$\{E^\ast_i(\cdot),n^\ast_i\}$, supposed it is then straightforward
to confirm that via the Strong Law of Large Numbers the empirical
distribution function of the samples $\bm{\Theta}_k$ drawn under INS,
$F^\mathrm{INS}(\bm{\Theta})$, will converge in distribution to this
asymptotic form; the convergence-determining classes here being simply
the (lower open, upper closed) hyper-cubes in the compact subset,
$[0,1]^N$, of $\mathbb{R}^N$.  From
$F^\mathrm{INS}(\bm{\Theta}) \stackrel{\text{d}}{\hbox{$\Rightarrow$}}\
F^\ast(\bm{\Theta})$ we have 
\begin{equation}
g^{[\mathrm{biased}]}(\bm{\Theta}), g^{[\mathrm{unbiased}]}(\bm{\Theta}) \rightarrow \frac{\partial^N}{\partial \Theta_1,\ldots,\partial \Theta_N}F^\ast(\bm{\Theta}),
\end{equation}
and thus
\begin{equation}
\mathrm{E}[\hat{\mathcal{Z}}^\mathrm{INS}] = \int\frac{f(\bm{\Theta})g^{[\mathrm{biased}]}(\bm{\Theta})}{g^{[\mathrm{unbiased}]}(\bm{\Theta})}d\bm{\Theta} \rightarrow \int f(\bm{\Theta}) d\bm{\Theta} = \mathcal{Z},
\end{equation}
and hence (with
the corresponding $\mathrm{Var}[\hat{\mathcal{Z}}^\mathrm{INS}]\rightarrow 0$) the
consistency of $\hat{\mathcal{Z}}^\mathrm{INS}$.\\

\subsection{Variance breakdown of INS} 

Given the evident dependence of
the INS variance on three distinct sources---\textsc{(i)} the
stochasticity of the live point set, and its decompositions,
$\{E_i(\cdot)\}$; \textsc{(ii)} the negative binomial sampling of the
$n_i$; and \textsc{(iii)} the importance sampling variance of the
drawn $f(\bm{\Theta}_k)/g(\bm{\Theta}_k)$---it makes good sense to
break these components down into their contributory terms via the Law
of Total Variance as follows:
\begin{eqnarray}
\mathrm{Var}_{\pi(\{E_i(\cdot)\},\{n_i\},\{\bm{\Theta}_k\})}[\hat{\mathcal{Z}}^\mathrm{INS}] \nonumber \\
&& \hspace*{-4cm} = \mathrm{Var}_{\pi(\{E_i(\cdot)\})}[\mathrm{E}_{\pi(\{n_i\},\{\bm{\Theta}_k\}|\{E_i(\cdot)\})}[\hat{\mathcal{Z}}^\mathrm{INS}]]\ \nonumber \\
&& \hspace*{-4cm} + \mathrm{E}_{\pi(\{E_i(\cdot)\})}[\mathrm{E}_{\pi(\{\bm{\Theta}_k\}|\{E_i(\cdot)\})}[\mathrm{Var}_{\pi(\{n_i\}|\{\bm{\Theta}_k\},\{E_i(\cdot)\})} [
\hat{\mathcal{Z}}^\mathrm{INS} ]]] \nonumber  \\ 
&& \hspace*{-4cm} + \mathrm{E}_{\pi(\{E_i(\cdot)\})}[\mathrm{Var}_{\pi(\{\bm{\Theta}_k\}|\{E_i(\cdot)\})}[
\mathrm{E}_{\pi(\{n_i\}|\{\bm{\Theta}_k\},\{E_i(\cdot)\})} [
\hat{\mathcal{Z}}^\mathrm{INS} ]]].
\end{eqnarray}

Now the first term represents explicitly the variance contribution from the
inherent randomness of the ellipsoidal decomposition sequence,
$\{E_i(\cdot)\}$, which we might suppose negligble provided the geometric
exploration of the posterior has been `sufficiently thorough', meaning
that the $N_\mathrm{live}$ point set has evolved through all the
significant posterior modes.  The second and third terms represent the
negative binomial sampling and `ordinary' importance sampling
variance contributions expected under the distribution of
$\{E_i(\cdot)\}$.  With the realised $\{E_i(\cdot)\}$ being, of course,
an unbiased draw from its parent distribution any unbiased estimator of these two additional variance components
 applied to our realised $\{n_i\}$ and $\{\bm{\Theta}_k\}$ could be
 considered likewise unbiased.  However, no such estimators are
 readily available, hence we pragmatically suppose the second
 term also negligble and make do with the `ordinary'
 importance sampling estimator, given by Eq.~(\ref{eq:IS-var-Z}), for
 the third term.  

Acknowledging the possibility for under-estimation of the INS variance
 in this way it becomes prudent to consider strategies for minimising
 our unaccounted variance contributions.  The first, suggested by our
 preceeding discussion of asymptotic consistency for the INS, is to
 maximise the size of the live point set used.  Of course, whether for
 INS or vanilla NS with {\sc MultiNest} we have no prescription for
 the requisite $N_\mathrm{live}$, and the range of feasible
 $N_\mathrm{live}$ will often be strongly limited by the available
 computational resources.  Hence we can give here only the general
 advice of cautious application; in particular it may be best to
 confirm a reasonable similarly between the estimated variance from
 Eq.~(\ref{eq:IS-var-Z}) above and that observed from repeat
 simulation at an $N_\mathrm{live}$ of manageable runtime prior to
 launching {\sc MultiNest} at a more expensive $N_\mathrm{live}$.  The
 second means of reducing the variance in our two unaccounted terms is
 to stick with the default mode of {\sc MultiNest}, rather than opt
 for `constant efficiency mode', since by bounding the maximum rate at
 which the ellipsoidal decompositions may shrink towards the posterior
 mode we automatically reduce the variance in the random variable,
 $\{E_i(\cdot)\}$, \textit{and} that of $\{n_i\}$ and
 $\hat{\mathcal{Z}}$ conditional upon it!

\section{Some Measure-Theoretic Considerations}

When outlining in Sec.\ \ref{sec:multinest} the transformation from
integration of $\mathcal{L}(\bm{\Theta})$ over
$\pi(\bm{\Theta})d\bm{\Theta}$ to integration of $\mathcal{L}(X)$ over
$dX$ (the prior mass cumulant) at the heart of the NS algorithm, we
elected, in the interest of simplicity, to omit a number of underlying
measure-theoretic details.  The significance of these are perhaps only
of particular relevance to understanding the use of the NS posterior
weights, $\mathcal{L}_i w_i / \hat{\mathcal{Z}}$ from
Eq.~(\ref{eq:12}), for inference regarding functions of $\bm{\Theta}$
(e.g.\ its first, second, and higher-order moments) with respect to
the posterior density.  However, as this issue has been raised
by \citet{cho10} and we feel that their Lemma 1 deserves clarification
we give here a brief measure-theoretic formulation of NS with this in
mind.

As with many Bayesian inference problems we begin by supposing the
existence of two well-defined probability spaces: \textsc{(i)} that of
the prior with (normalised) probability measure, $P_\pi$, defined for
events in the $\sigma$-algebra, $\Sigma_{\bm{\Theta}}$, of its sample
space, $\Omega_{\bm{\Theta}}$, i.e.,
$(P_\pi,\Omega_{\bm{\Theta}},\Sigma_{\bm{\Theta}})$, and \textsc{(ii)}
that of the posterior with measure $P_{\pi^\prime}$ defined on the
same space, i.e.,
$(P_{\pi^\prime},\Omega_{\bm{\Theta}},\Sigma_{\bm{\Theta}})$.
Moreover, we suppose that each may be characterised by its
Radon--Nikodym derivative with respect to a common $\sigma$-finite
baseline measure on a complete, separable metric space; that is,
$P_\pi(A \in \Sigma_{\bm{\Theta}}) = \int_A
\pi(\bm{\Theta}) \{d\bm{\Theta}\}$ and $P_{\pi^\prime}(A \in \Sigma_{\bm{\Theta}}) = \int_A
\pi(\bm{\Theta}) \mathcal{L}(\bm{\Theta})/\mathcal{Z}
\{d\bm{\Theta}\}$.  NS then proposes to construct the induced measure, $P_X$,
on the $\sigma$-algebra, $\Sigma_X$, generated by the Borel sets of the sample space, $\Omega_X = [0,1]
\in \mathbb{R}$, and defined by the transformation, $X:\Omega_{\bm{\Theta}}
\mapsto\Omega_X$ with $X(\bm{\Theta}^\prime) = \int_{\{\bm{\Theta} :
  \mathcal{L}(\bm{\Theta}) > \mathcal{L}(\bm{\Theta}^\prime)\}}
\pi(\bm{\Theta})\{ d\bm{\Theta}\}$.  The validity of which
requires only the measurability of this transformation (i.e., $X^{-1}B
\in \Sigma_{\bm{\Theta}}$ for all $B \in \Sigma_{X}$); e.g.\ in the
metric space $\mathbb{R}^k$ with reference Lebesgue measure, the almost
everywhere continuity of $\mathcal{L}(\bm{\Theta})$.  However, for the
proposed Riemann integration of vanilla NS to be valid we will also
need the induced $P_X$ to be absolutely continuous with respect to
the Lebesgue reference measure on $[0,1]$, such that we can write $P_{X}(B \in \Sigma_{X}) = \int_B
\mathcal{L}(X)/\mathcal{Z}
\{d\bm{X}\}$.  The additional condition for this given by
\citet{cho10} is that $\mathcal{L}(\bm{\Theta})$
has connected support.  To state the objective of vanilla NS in a
single line: if we can compute $\mathcal{L}(X)$ we can find $\mathcal{Z}$ simply by
solving for $\int_0^1 \mathcal{L}(X)/\mathcal{Z} \{dX\} = 1$.

In their Sec.\ 2.2 \citet{cho10} examine the NS importance sampling
estimator proposed for the posterior expectation of a given function
$f(\bm{\Theta})$, namely \begin{equation}
\mathrm{E}_{\pi^\prime}[f(\bm{\Theta})] = \int_{\Omega_{\bm{\Theta}}}
f(\bm{\Theta}) \pi(\bm{\Theta}) \mathcal{L}(\bm{\Theta}) /\mathcal{Z}
\{d\bm{\Theta}\}, \end{equation}
which one may approximate with $\hat{\mathrm{E}}[f(\bm{\Theta})]= \sum f(\bm{\Theta}_i)\mathcal{L}_i w_i$ from 
  NS Monte Carlo sampling.  They note that $f(\bm{\Theta})$ is
in this context a noisy estimator of $\tilde{f}(X) =
\mathrm{E}_{\pi}[f(\bm{\Theta}) | \mathcal{L}(\bm{\Theta}) = \mathcal{L}(X)]$,
and propose in their Lemma 1 that the equality, \begin{equation} \int_0^1
\tilde{f}(X)  \mathcal{L}(X) \{dX\} = \int_{\Omega_{\bm{\Theta}}}
f(\bm{\Theta}) \pi(\bm{\Theta}) \mathcal{L}(\bm{\Theta})
\{d\bm{\Theta}\},\label{eqnx} \end{equation} holds \textit{when $\tilde{f}(X)$ is absolutely continuous}.
We agree that this is true and \citet{cho10} give a valid proof in
their Appendix
based on the Monotone Convergence Theorem.  However, we suggest that
given the already supposed validity of the measure $P_X$, and its
Radon--Nikodym derivative with respect to the reference Lebesgue measure,
$\{dX\}$, upon which
NS is based, the equality of the above relation is already true without
absolute continuity via the change of variables theorem \citep{hal50}, in the sense
that wherever one side exists the other exists and will be equal to
it.  One trivial example of a discontinuous $\tilde{f}(X)$ for which
both sides of \ref{eqnx} exist and are equal is that induced by the indicator
function for $\mathcal{L}(\bm{\Theta}) > X^\ast$.  To see that the
$\tilde{f}(X)$ corresponding to a given, measurable $f(\bm{\Theta})$ has  
the stated interpretation as a conditional expectation we observe that
$\mathrm{E}_{\pi}[f(\bm{\Theta}) | \mathcal{L}(\bm{\Theta}) = \mathcal{L}(X)]$ may be written as
\begin{equation}
\int_{\Omega_{\bm{\Theta}}} f(\bm{\Theta}) \pi(\bm{\Theta}) \mathrm{I}_{\mathcal{L}(\bm{\Theta})=\mathcal{L}(X)}\{d\bm{\Theta}\},
\end{equation}
a function of $X$, using the interpretation of conditional
distributions as derivatives \citep{pfa79}.  Thus,
\begin{equation}
\int_{\Omega_{\bm{\Theta}}} f(\bm{\Theta}) \pi(\bm{\Theta})
\mathcal{L}(\bm{\Theta}) \{d\bm{\Theta}\} = \int_0^1 \left(
\int_{\Omega_{\bm{\Theta}}} f(\bm{\Theta})
\pi(\bm{\Theta}) \mathrm{I}_{\mathcal{L}(\bm{\Theta})=\mathcal{L}(X)}
 \{d\bm{\Theta}\} \right) \mathcal{L}(X) \{dX\}.
\end{equation}

\end{appendices}

\bibliographystyle{ECA_jasa}
\bibliography{references}

\label{lastpage}
\end{document}